\documentclass[aps,prd,twocolumn,amsmath,amssymb,showpacs,floatfix,superscriptaddress,nofootinbib]{revtex4-1}

\pdfoutput=1
\usepackage{natbib}
\usepackage{amsmath}
\usepackage{amssymb}
\usepackage{latexsym}
\usepackage{bm}   
\usepackage{graphicx,epstopdf}
\usepackage{epstopdf}
\DeclareGraphicsExtensions{.ps, .png}
\usepackage{dcolumn} 
\usepackage{multirow}
\usepackage{appendix}
\usepackage{footnote}
\usepackage{tabularx,ragged2e,booktabs}
\usepackage{breqn}
\usepackage[normalem]{ulem}
\usepackage{float}
\restylefloat{table}

\newcolumntype{C}[1]{$>${\Centering}m{#1}}

\newcommand{\al}[1]{\begin{align} #1 \end{align}}

\newcommand{\vk}{\boldsymbol{k}}

\def\dd{\mathrm{d}}

\newcommand{\kk}{^{\kappa\kappa}(l)}

\newcommand{\gk}{^{g \kappa}(l)}

\newcommand{\kkk}{\kappa \kappa \kappa}
\newcommand{\kkg}{\kappa \kappa g}
\newcommand{\kgg}{\kappa gg}

\newcommand{\beq}{\begin{equation}}
\newcommand{\eeq}{\end{equation}}
\newcommand{\bea}{\begin{eqnarray}}
\newcommand{\eea}{\end{eqnarray}}

\newcommand{\skaco}[1]{\langle{#1}\rangle}
\def\bk{\bm{k}}

\usepackage{color}
\definecolor{ultramarine}{rgb}{0.07, 0.04, 0.56}
\definecolor{cadmiumgreen}{rgb}{0.0, 0.42, 0.24}
\definecolor{indigo(dye)}{rgb}{0.0, 0.25, 0.42}
\usepackage[linktocpage=true]{hyperref}
\hypersetup{
colorlinks=true,
citecolor=ultramarine,
linkcolor=cadmiumgreen,
urlcolor=indigo(dye),
pdfauthor={},
pdftitle={},
pdfsubject={}
}

\def\lsim{\mathrel{\raise.3ex\hbox{$$<$$\kern-.75em\lower1ex\hbox{$\sim$}}}}
\def\gsim{\mathrel{\raise.3ex\hbox{$$>$$\kern-.75em\lower1ex\hbox{$\sim$}}}}

\definecolor{darkgreen}{cmyk}{0.85,0.2,1.00,0.2} 
\definecolor{purple}{cmyk}{0.5,1.0,0,0}

\begin{document}
	
\title{Cross-bispectra Constraints on Modified Gravity Theories from Nancy Grace Roman Space Telescope and Rubin Observatory Legacy Survey of Space and Time}

\author{Chen Heinrich}\email{chenhe@caltech.edu}
\affiliation{$Jet\ Propulsion\ Laboratory,\ California\ Institute\ of\ Technology,\ Pasadena,\ California\ 91109,\ USA$}
\affiliation{$California\ Institute\ of\ Technology,\ Pasadena,\ California\ 91109,\ USA$}

\author{Olivier Dor\'{e}}
\affiliation{$Jet\ Propulsion\ Laboratory,\ California\ Institute\ of\ Technology,\ Pasadena,\ California\ 91109,\ USA$}
\affiliation{$California\ Institute\ of\ Technology,\ Pasadena,\ California\ 91109,\ USA$}

\begin{abstract}

One major goal of upcoming large-scale-structure  surveys is to constrain dark energy and modified gravity theories. In particular, galaxy clustering and gravitational lensing convergence are probes sensitive to modifications of general relativity. While the standard analysis for these surveys typically includes power spectra or 2-point correlation functions, it is known that the bispectrum contains more information, and could offer improved constraints on parameters when combined with the power spectra. However, the use of bispectra has been limited so far to one single probe, e.g. the lensing convergence bispectrum or the galaxy bispectrum. In this paper, we extend the formalism to explore the power of cross-bispectra between different probes, and exploit their ability to break parameter degeneracies and improve constraints. We study this on a test case of lensing convergence and galaxy density auto- and cross-bispectra, for a particular sub-class of Horndeski theories parametrized by $c_M$ and $c_B$. Using the 2000 deg$^2$ notional survey of the Nancy Grace Roman Space Telescope with overlapping photometry from the Rubin Observatory Legacy Survey of Space and Time, we find that a joint power spectra and bispectra analysis with three redshift bins at $l_{\rm max} = 1000$ yields $\sigma_{c_M} = 1.0$ and $\sigma_{c_B} = 0.3$, both a factor of $\sim$1.2 better than the power spectra results; this would be further improved to $\sigma_{c_M} = 0.7$ and $\sigma_{c_B} = 0.2$ if $l_{\rm max} = 3000$ is taken. Furthermore, we find that using all possible cross-bispectra between the two probes in different tomographic bins improves upon auto-bispectra results by a factor of 1.3 in $\sigma_{c_M}$, 1.1 in $\sigma_{c_B}$ and 1.3 in $\sigma_{\Omega_m}$. We expect that similar benefits of using cross-bispectra between probes could apply to other science cases and surveys.

\end{abstract}

\maketitle

\section{Introduction}
\label{sec:intro}

Recent measurements have shown that the Universe is consistent with a $\Lambda$CDM model, exhibiting an epoch of accelerated expansion of the Universe. The observation of the accelerated epoch poses some theoretical challenges. Three possibilities are typically considered: 1) a cosmological constant; 2) a scalar field giving rise to a possibly evolving energy density, dubbed dark energy; 3) that the theory of general relativity (GR) does not hold on cosmological scales, requiring modified gravity theories (MG). While GR is a well-tested theory on some scales such as the solar system scale, whether it can be successfully extrapolated to all cosmological scales over many orders of magnitude is still an assumption that remains to be tested. Cosmology holds great promise for testing alternative theories of gravity as they would leave distinguishable signatures on probes like the clustering of galaxies, gravitational lensing, redshift space distortions among many others.

Upcoming Stage-IV large-scale-structure surveys such as 
EUCLID\footnote{\url{www.euclid-ec.org}}~\cite{2011arXiv1110.3193L},
%
Nancy Grace Roman Space Telescope\footnote{\url{https://wfirst.gsfc.nasa.gov/}}~\cite{Spergel:2013tha},
Rubin Observatory Legacy Survey of Space and Time (LSST)\footnote{\url{https://www.lsst.org/}}~\cite{Ivezic:2008fe}
and DESI\footnote{\url{http://desi.lbl.gov/}}~\cite{Aghamousa:2016zmz} 
are optimally designed to maximize constraints on dark energy and modified gravity. By detecting hundreds of millions of galaxies in large areas of the sky in imaging and spectroscopy, these surveys allow us to construct powerful statistical probes that can distinguish between different theories. the Roman Space Telescope among others, is a particularly promising survey that would notionally map a 2,000 deg$^2$ area of the sky in depth in both spectroscopy and imaging, enabling multi-probe analysis as well as exquisite systematics control
~\cite{Spergel:2013tha, Eifler:2020vvg, Eifler:2020hoy, 2019ApJ...877..117H}.

To take advantage of these superb capabilities of future surveys, it is no longer sufficient to restrict ourselves to the typical analysis using power spectra or 2-pt correlation functions. More information is known to exist in higher-order statistics such as the bispectrum. Combining bispectrum with power spectrum observations typically lead to improved constraints on parameters. In this paper, we explore how adding bispectra, and in particular cross-bispectra measurements between two probes, provides improved constraints on modified gravity theories for the overlapping 2,000 deg$^2$ between the Roman Space Telescope and the LSST survey.

Just like using cross-power spectra can break parameter degeneracies leading to improved constraints, we expect similar effect by cross-correlating probes in the bispectra. Typically the lensing convergence bispectrum or the galaxy bispectrum have been studied individually (e.g. \cite{Takada:2003ef, Rizzato:2018whp, Yankelevich:2018uaz, Tellarini:2016sgp}). Here, we combine for the first time the galaxy and lensing convergence auto-bispectra as well as their cross-bispectra between these probes in different tomographic bins, to exploit their potential to improve on parameter constraints. We demonstrate such improvement on a particular subclass of Horndeski models for the Roman Space Telescope, but expect similar benefits to extend potentially to other science cases or other surveys. 

The paper is structured as follows. In section~\ref{sec:MG}, we present the background on Horndeski theories, and define the subclass of MG models that we study. In section~\ref{sec:ps} and~\ref{sec:bis} respectively, we describe the modeling of the power spectrum and bispectrum observables, as well as the effects of modified gravity on them. We present the Fisher forecast formalism in Section~\ref{sec:fisher} used to obtain the results in Section~\ref{sec:results}, which we summarize and discuss in Section~\ref{sec:conclusion}.  
\begin{widetext}
\section{Horndeski Theory}
\label{sec:MG}

We now describe the subclass of Horndeski theories studied in this paper. Horndeski theory~\cite{Horndeski:1974wa} is the most general theory of gravity in four dimensions postulating a scalar field in addition to the metric tensor, while giving rise to second order equations of motion. Its Lagrangian 
\begin{equation}
S[g_{\mu\nu},\phi]=\int\mathrm{d}^{4}x\,\sqrt{-g}\left[\sum_{i=2}^{5}\frac{1}{8\pi G_{\text{N}}}{\cal L}_{i}[g_{\mu\nu},\phi]\,+\mathcal{L}_{\text{m}}[g_{\mu\nu},\psi_{M}]\right]\,,\label{eq:action}
\end{equation}
contains four arbitrary functions $\{G_a(\phi, X)$, $a = 2, 3, 4, 5\}$ for the scalar field $\phi$
\begin{eqnarray}
{\cal L}_{2} & = & G_{2}(\phi,\,X)\,,\label{eq:L2}\\
{\cal L}_{3} & = & -G_{3}(\phi,\,X)\Box\phi\,,\label{eq:L3}\\
{\cal L}_{4} & = & G_{4}(\phi,\,X)R+G_{4X}(\phi,\,X)\left[\left(\Box\phi\right)^{2}-\phi_{;\mu\nu}\phi^{;\mu\nu}\right]\,,\label{eq:L4}\\
{\cal L}_{5} & = & G_{5}(\phi,\,X)G_{\mu\nu}\phi^{;\mu\nu}-\frac{1}{6}G_{5X}(\phi,\,X)\left[\left(\Box\phi\right)^{3}+2{\phi_{;\mu}}^{\nu}{\phi_{;\nu}}^{\alpha}{\phi_{;\alpha}}^{\mu}-3\phi_{;\mu\nu}\phi^{;\mu\nu}\Box\phi\right]\,,\label{eq:L5}
\end{eqnarray}
 where $X \equiv -\partial_{\mu}\phi \delta^{\mu} \phi$, $G_{\mu\nu}$ is the Einstein tensor and $R$ is the Ricci scalar. The matter Lagrangian is denoted by $\mathcal{L}_{\text{m}}$ in Eq.~\ref{eq:action}, where $g_{\mu\nu}$ is the Jordan-frame metric, $\psi_{M}$ are the matter fields and $G_N$ is the Newton gravitational constant. The partial derivatives are denoted with subscripts $\phi, X$, e.g. $G_{5X} \equiv \partial G_5/\partial X$; the covariant derivatives are denoted with subscript $;$. 

\end{widetext}

An alternative and physically more meaningful basis of functions can be obtained by expanding the action to second order in linear perturbations of $g_{\mu\nu}$, $\phi$ and other matter fields~\cite{Bellini:2014fua}. The action would consist of terms quadratic in the perturbations, each multiplied by time dependent functions which are only affected by the background cosmology, so that once the background expansion is fixed, the modifications to Einstein's equations in Horndeski theories are specified by four functions of time. A set of basis for these functions with direct physical interpretations was identified in Ref.~\cite{Bellini:2014fua} as  $\alpha_i$, $i \in \{M, T, B, K$\}:
\begin{enumerate}
\item{$\alpha_M \equiv \mathrm{d\, ln} M_*^2/\mathrm{d\,ln} a$, the running of the Planck mass, controls the strength of gravity given the initial value of $M_*^2$;}
\item{$\alpha_T \equiv c_T^2 - 1$, the tensor speed excess, controls the excess speed of the gravitational waves propagation with respect to light;}
\item{$\alpha_B$, the braiding, parametrizes the mixing between the scalar field and metric kinetic terms~\cite{Bettoni:2015wta};}

\item{$\alpha_K$, the kineticity, is the coefficient of the kinetic term for the scalar d.o.f. before demixing~\cite{Bellini:2014fua}. }
\end{enumerate}

The full form of the $\alpha$'s can be found in the Appendix A.3 of Ref.~\cite{Zumalacarregui:2016pph} and further explanations on these parameters can be found, e.g. in Ref.~\cite{Bellini:2014fua} and references therein. 

Because of the gravitational waves observations of GW170817, $\alpha_T$ has been highly constrained~\cite{Ezquiaga:2017ekz, Kase:2018aps}, so we will fix it to practically zero throughout our work. We also cannot constrain $\alpha_K$ with sub-horizon probes we are using in this work. Because $\alpha_K$ increases the relative strength of kinetic to gradient term, it lowers the sound speed and hence the sound horizon to below the cosmological horizon, where a quasi-static configuration is reached~\cite{Bellini:2014fua, Alonso:2016suf}, so $\alpha_K$ cannot be constrained with the quasi-static scales (although it can be probed with ultra-large scales~\cite{Gleyzes:2015rua}). As a result, we will focus solely on constraining $\alpha_M$ and $\alpha_B$ in this work .

We choose to restrict ourselves, following Ref.~\cite{Yamauchi:2017ibz}, to a class of models $\{\alpha_i\}$ whose time dependence follows the time evolution of the dark energy density:
\beq
\alpha_{i}(a) = c_{i} \Omega_{\mathrm{DE}}(a),
\eeq
where $a$ is the scale factor and $c_{i}$'s are constants of proportionality. 
Note that this parametrization is purely phenomenologically motivated, and may imply fine tuning between terms at the Lagrangian level. 

We further restrict ourselves to consider matter that is minimally coupled to the metric without direct couplings to the scalar, so the effect of the modified gravity sector on matter is only mediated through the gravitational potential like in the case of general relativity. Finally, the background expansion is fixed to that of $\Lambda$CDM.

To compute the matter power spectrum of the considered models and the evolution of various quantities, we use the public code \texttt{hi\_class}\footnote{hi\_class: \url{http://miguelzuma.github.io/hi_class_public/} }~\cite{Zumalacarregui:2016pph}: Horndeski in CLASS (Cosmic Linear Anisotropy Solving System~\cite{2011arXiv1104.2932L}). Our choice of parametrization corresponds exactly to the \texttt{propto\_omega} option in \texttt{hi\_class} that takes $c_i$'s as input parameters, as well as the initial $M_{*}$ which we set to 1 in units of the Planck mass $M_{\rm pl}$ to match GR solutions at early times.

\section{Power Spectrum Observables}
\label{sec:ps}

We now describe the power spectrum observables (galaxy clustering, lensing convergence and their cross-power) and how we obtain them given the linear matter power spectrum $P_m(k)$ from \texttt{hi\_class}. We first define the observables in section~\ref{sec:ps_def} and describe their modeling in GR, then we introduce the modifications from the  Horndeski theory parameters in section~\ref{sec:ps_mg} and study their impacts on the power spectra.

\subsection{Definitions}
\label{sec:ps_def}

A projected observable in two-dimensions $X(\bm{\theta})$ can be described in terms of its Fourier transform $X_{\bm{l}}$
\beq
X(\bm{\theta})=\sum_{\bm{l}}
X_{\bm{l}} e^{i\bm{l} \cdot \bm{\theta}}, 
\eeq
where $\bm{l}$ is the Fourier wavevector.
The angular power spectrum $C^{XY}(l)$ is defined as
\begin{eqnarray}
\skaco{X_{\bm{l}}Y_{\bm{l'}}}
&=&(2\pi)^2\delta^D(\bm{l}+\bm{l'})C^{XY}(l),
\end{eqnarray}
where $\skaco{\cdots}$ denotes the ensemble average and $\delta^D(\bm{l})$ is the Dirac delta function.

In the Limber approximation, the angular power spectrum between two probes $X$ and $Y$ is given by
\beq
C^{XY}(l) = \int d\chi W^X (\chi) W^Y (\chi) \chi^{-2} P_m(k = l/\chi; \chi),
\label{eq:cldef}
\eeq
where $P_m(k; \chi)$ is the three-dimensional matter power spectrum, $\chi$ is the comoving distance and $W^{X}$ the kernel for the probe $X$. In this work, we consider $X, Y\in \{\kappa, g\}$, where $\kappa$ is lensing convergence and $g$ is the galaxy density constrast.

Two galaxy samples are involved, the source sample which contain the background galaxies being lensed, and the lens sample which contains galaxies that act as a lens to the background galaxies. For our galaxy sample for $g$, we use the lens sample. We do not use a spectroscopic galaxy sample, though it is also possible to do so. 


For galaxy density contrast $g$ we have
\beq
W^g(\chi) \equiv b_{1}(\chi) \frac{p^{\rm len}(z(\chi))}{\bar{n}^{\rm len}} \frac{dz}{d\chi},
\eeq
where $b_{1}$ is the linear galaxy bias at $\chi$, 
$p^{\rm len}$ and $\bar{n}^{\rm len}$ are the redshift distribution and the average number density of the lens galaxy sample respectively.

For the lensing convergence $\kappa$ we have
\beq
\label{eq:Wk}
W^{\kappa}(\chi) = 
\frac{3\Omega_{m,0} H_0^2}{2  c^2 a(\chi) \bar{n}^{\rm src}} 
\int_{\chi}^{\chi_H} d\chi_s p^{\rm src}(z(\chi_s)) \frac{dz}{d\chi_s} \frac{\chi (\chi_s - \chi)}{\chi_s},
\eeq
where $p^{\rm src}$ and $\bar{n}^{\rm src}$ are the redshift distribution and the average number density of the source galaxy sample respectively, $\Omega_{m,0}$ is the matter density at $z = 0$, $H_0$ is the Hubble constant today, and $a$ is the scale factor.

We also consider tomography since the redshift dependence of the lensing kernel gives additional information that generally improves constraints. The power spectrum between $X$ in redshift bin $i$ and $Y$ in redshift bin $j$ is
\beq
C_{(ij)}^{XY}(l) = \int d\chi W^X_{(i)}(\chi) W^Y_{(j)}(\chi) \chi^{-2} P_m(k = l/\chi; \chi),
\label{eq:clijdef}
\eeq
where
\beq
\label{eq:Wg_i}
W^g_{(i)}(\chi) \equiv b_{1,(i)}(\chi) \frac{p_{(i)}^{\rm lens}(z(\chi))}{\bar{n}_{(i)}^{\rm lens}} \frac{dz}{d\chi},
\eeq
where $b_{1,(i)}$ is the linear galaxy bias in the redshift bin $i$.
and
\bea
\label{eq:Wk_i}
W_{(i)}^{\kappa}(\chi)&=& \frac{3\Omega_{m,0} H_0^2}{2 c^2  a(\chi)\bar{n}_{(i)}^{\rm src}} 
\int_{{\rm max}\{\chi,\chi_{i}\}}^{\chi_{i+1}} \!\!d\chi_{s}~ 
p^{\rm src}_{(i)}(z(\chi_s))  \notag \\ 
&\times & \frac{dz}{d\chi_s} \frac{\chi (\chi_{\rm s}-\chi) }{\chi_s}, \;\;\;\;
\mathrm{for\;} \chi>\chi_{i+1},
\eea
for $\chi\le\chi_{i+1}$ and 
\beq
W_{(i)}^{\kappa}(\chi) = 0, \mathrm{for\;} \chi>\chi_{i+1}.
\eeq
The quantities $p_{(i)}$ and $\bar{n}_{(i)}$ are now the redshift distribution and the average number density respectively in redshift bin $i$ of the corresponding sample.

In principle, the galaxy bias is also a function of redshift and can be modeled, but we have chosen to model it as a nuisance parameter that varies with the redshift bin. So for three redshift bins, there are three values of $b_{1,(i)}$ to be marginalized over in the Fisher analysis, whereas for one redshift bin, there is only one. The fiducial values of $b_{1,(i)}$ are fixed at 1. 

Note that for both the lens and source populations, we have ignored errors in the photometric redshifts for simplicity which is reasonable given that we consider only few large redshift bins. 

Now the observed power spectrum 
has additional noise contributions
\beq
\tilde{C}_{(ij)}^{XY}(l)=C_{(ij)}^{XY}(l) + N_{(ij)}^{XY},
\label{eq:obscl}
\eeq
where
\beq
N_{(ij)}^{gg} = \delta_{ij}\frac{1}{\bar{n}_{(i)}^{\rm lens}}
\label{eq:ngg}
\eeq
is the shot noise from the Poisson sampling of the underlying matter density for galaxies, and
\beq
N_{(ij)}^{\kappa\kappa} = \delta_{ij}\frac{\sigma_{\epsilon}^2}{\bar{n}_{(i)}^{\rm src}},
\label{eq:n}
\eeq
accounts for the noise in the lensing convergence power spectrum due to the intrinsic ellipticity of the source galaxies. Furthermore, we assume $N_{(ij)}^{g\kappa} = 0$ here for simplicity. In reality, there would be additional correlations that arise from systematics effects such as intrinsic alignments. These would impact the constraints on cosmological parameters (see e.g. Ref.~\cite{Eifler:2020vvg}), with a degree that depends on how our understanding of systematics evolve in the next decade; we leave studying those effects to a future paper. 

For the particular Roman + LSST survey configuration, we start with the same distributions adopted in Ref.~\cite{Eifler:2020vvg}, which follows Ref.~\cite{2019ApJ...877..117H} in applying the Roman exposure time calculator~\cite{2012arXiv1204.5151H} on the CANDELS data set -- the detailed procedure may be found in section 2.1 of Ref.~\cite{Eifler:2020vvg} under item ``Define the galaxy samples". Here we use slightly different total number densities $\bar{n}
^{\rm len} = \bar{n}
^{\rm src} = 50$ (c.f. $\bar{n}^{\rm len} = 66$ and $\bar{n}^{\rm src} = 51$), but we do not expect our results to be significantly impacted as they are dominated by the lensing convergence whose noise spectrum, controlled by $\bar{n}^{\rm src}$, is not significantly changed.

\begin{figure}[t]
\includegraphics[width=0.40\textwidth]{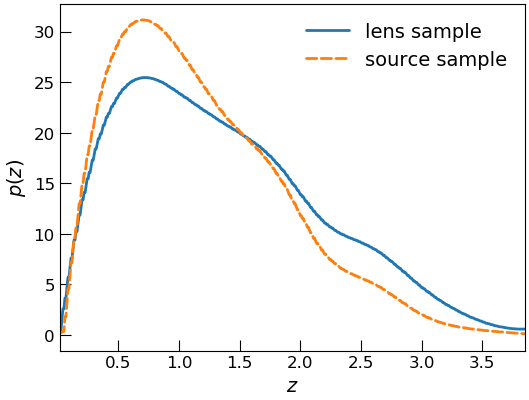}
\caption{Redshift distributions $p_{(i)}(z)$ for the overlapping 2000~deg$^2$ survey from Roman and LSST for the lens galaxy sample (blue solid) and source galaxy sample (orange dashed).
}
\label{fig:dndz}
\end{figure}

In Fig.~\ref{fig:dndz} we show the redshift distributions $p(z)$ for the lens sample (blue solid) and the source samples (orange dashed) used in this paper over the notional 2000~deg$^2$ overlapping survey between the Roman and LSST. As stated previously, we do not include photometric redshift errors (contrary to Ref.~\cite{Eifler:2020vvg}). Our boundaries for the redshift bins will be chosen such that the total number of lens galaxies inside each bin is the same, so that the galaxy noise spectra in Eq.~\ref{eq:ngg} are constant between the redshift bins. As a result, the lensing convergence power spectrum will have slightly non-constant noise spectra as the source sample distribution is slightly different from the lens sample distribution. 

We model the covariance between the observed power spectra as
\bea
\label{eq:covps}
&&{\rm Cov}[\tilde{C}_{(ij)}^{XY}(l),\tilde{C}_{(ab)}^{X'Y'}(l')] \notag \\
&=&\frac{\delta_{ll'}}{(2l+1) f_{\rm sky}} 
\left[\tilde{C}^{XX'}_{(ia)}(l) \tilde{C}_{(jb)}^{YY'}(l)
+\tilde{C}^{XY'}_{(ib)}(l)\tilde{C}_{(ja)}^{YX'}(l)\right],
\nonumber\\
\eea
where $i,j,a,b \in \{1 ... n_{\rm zbin}\}$ and $f_{\rm sky}$ is the fractional of the sky observed. For the overlapping 2000 deg$^2$ survey of Roman + LSST, we have $f_{\rm sky} = 0.0485$. 

There is in principle a connected four-point function term due to the non-Gaussianity of the matter field, giving rise to the non-Gaussian covariance term, as well as a super-sample covariance term due to the finite area of the survey.
In Ref.~\cite{Rizzato:2018whp}, it was shown for the lensing power spectrum (which dominates contraints in this study), that including non-Gaussian and super-sample covariance could reduce the signal-to-noise ratio of the $C_l^{\kappa\kappa}$ by a factor of $2-3$ in the $l_{\rm max} = 1000 - 3000$ range without tomography. However, when tomography is used, the reduction becomes less than a factor of 2. Furthermore, we recall that a factor of 2 in signal-to-noise corresponds to a much smaller change in the marginalized error of individual parameters. As shown in Ref. \cite{Takada:2008fn}, a factor of at most 2 in signal-to-noise in the context of a eight-parameter Fisher analysis resulted in only a 10\% change on the individual parameter constraints. This is because if the volume of the Fisher ellipsoid in higher dimensional space was to be shrunk by half and uniformly in all directions, then each of the $N$ parameters would see their marginalized constraint change by a factor of $2^{1/N}$. So we will ignore non-Gaussian lensing contributions to the power spectrum covariance in this paper.

\subsection{Power Spectrum in Modified Gravity Theories}
\label{sec:ps_mg}

The impact of modified gravity on the power spectrum observables can mainly be parametrized by two phenomenological effects $\mu$ and $\Sigma$ in the quasi-static approximation, valid on scales much smaller than the cosmological horizon $H/k \ll 1$ and where the time derivatives of the perturbations are negligible compared to the spatial derivatives.
In these limits, the quantity $\mu$ parametrizes the strength of the 
effective gravitational coupling $G_{\rm eff}$ in units of 
the Newton constant $G_N=1/(8\pi M_{\rm pl}^2)$:
\beq
\mu=\frac{G_{\rm eff}}{G_N}\,,
\label{eq:mudef}
\eeq
which enters the modified Poisson equation that relates the gravitational potential $\Psi$ to the matter density contrast $\delta_m$
\beq
\frac{k^2}{a^2}\Psi =-4\pi G\mu \rho_m \delta_m.
\label{eq:poisson}
\eeq
Consequently the matter power spectrum is modified with a growth function $D(k,\chi)$ that unlike in GR, can now be in principle a scale-dependent function:
$P_m(k; \chi) = D(k; \chi)^2 P_m(k)$.

Now the gravitational slip parameter $\gamma$ is defined as
\beq
\gamma=\frac{\Phi}{\Psi}.
\label{eq:gamma}
\eeq
It follows from Eqs.~\ref{eq:poisson} and \ref{eq:gamma} that
\beq
\frac{k^2}{a^2}(\Phi + \Psi)=
8\pi G\,\Sigma\rho_m \delta_m\,, 
\label{eq:psieff}
\eeq
where 
\beq
\Sigma=\frac{1}{2}\mu (1+\gamma).
\label{eq:Sigma}
\eeq
The gravitational lensing is directly sensitive to $\Sigma$ because it probes the combination $\Phi + \Psi$: the lensing kernels of Eq.~\ref{eq:clijdef} are modified as
\bea
W^{\kappa}_{(i)}(\chi) \rightarrow \Sigma(k=l/\chi; \chi) W^{\kappa}_{(i)}(\chi),
\eea
in the Limber approximation and could also inherit in principle a scale-dependence through $\Sigma$. However, given our the Horndeski theory adopted here with phenomenologicallly parametrized $\alpha_i(a) \propto \Omega_{DE}(a)$, $\mu$ and $\Sigma$ are only time-dependent.

In quasi-static limit with no anisotropic stress and assuming pressureless matter and negligible velocity perturbation on subhorizon scales, we can relate $\mu$ and $\Sigma$ to $\alpha_i$ in Horndeski theories as follows\footnote{See also ``Notes on Horndeski Gravity" by Tessa Baker found at \url{http://www.tessabaker.space}}~\cite{Ishak:2019aay}: 
\beq
\label{eq:alpha2mu}
\mu = \frac{M_{\rm pl}^2}{M_*^2} \frac{\alpha c_s^2 (1+\alpha_T) + 2[-\alpha_B/(2 (1+\alpha_T)) + \alpha_T - \alpha_M]^2}{\alpha c_s^2},
\eeq
and
\beq
\label{eq:alpha2gamma}
\gamma = \frac{\alpha c_s^2 - \alpha_B[-\alpha_B/(2 (1+\alpha_T)) + \alpha_T - \alpha_M]}{\alpha c_s^2 (1+\alpha_T) + 2[-\alpha_B/(2 (1+\alpha_T)) + \alpha_T - \alpha_M]^2},
\eeq
where
\beq\label{eq:alpha}
\alpha = \alpha_K + \frac{3}{2} \alpha_B^2.
\eeq
Note that $\alpha_K$ ends up dropping out of the expression for $\mu$ and $\gamma$ as expected since its effects are not observable on quasi-static scales. We obtain the evolution of the quantities $M_*$ and $c_s$ from \texttt{hi\_class} and compute $\mu$ and $\Sigma$ using Eqs.~\ref{eq:Sigma}, ~\ref{eq:alpha2mu}$-$\ref{eq:alpha}. 

In the forecast work to follow, we actually fix the fiducial model to be not exactly but close to GR, to avoid the numerical singularity at $c_i = 0$. We adopt as fiducial MG parameters $\{c_B = c_M = c_T = 0.05, c_K = 0.1\}$. While non-zero $\alpha_M$ and $\alpha_T$ in the fiducial model means that a gravitational slip signal can be generated with a non-zero $c_B$ (otherwise not present), we have verified that lowering the fiducial values to $\{c_M = c_B = 0.025, c_T = c_K = 0.005\}$ actually produces only slightly more constraining results. So the above choice is still a conservative one. 

\begin{figure}[t]
\includegraphics[width=0.40\textwidth]{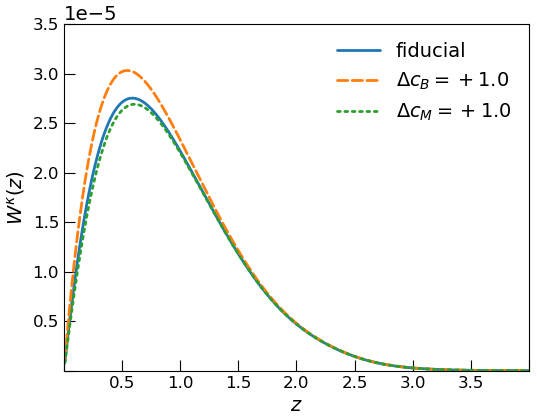}
\caption{Lensing kernel $W^{\kappa}(z)$ without tomography for the fiducial model (blue solid) and variations from it with $\Delta c_B = 1$ (orange dashed) and $\Delta c_M = 1$ (green dotted). The effects on the lensing kernel from $c_B$ and $c_M$ are opposite, resulting in opposite changes in the lensing-related power spectrum observables in Fig.~\ref{fig:cl}.}
\label{fig:wk}
\end{figure}

\begin{figure}[t]
\includegraphics[width=0.48\textwidth]{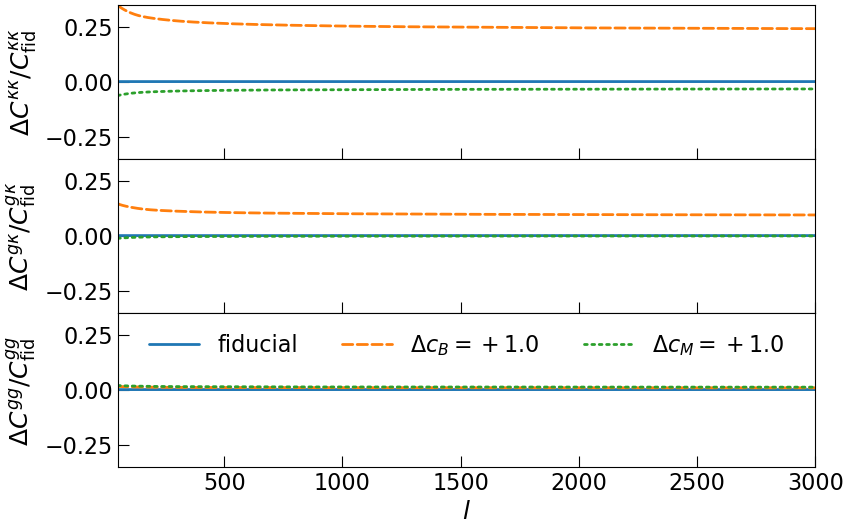}
\caption{Fractional deviation from the fiducial model for varying $\Delta c_B = 1$ (orange dashed) and $\Delta c_M = 1$ (green dotted) separately. The changes due to MG on the lensing probes $C\kk$ and $C\gk$ are dominated by the effects through the lensing kernel rather than the matter power spectrum; on the other hand, the galaxy clustering probe $C^{gg}(l)$ is only sensitive to the much smaller effects of modified growth in the power spectrum. Furthermore, the fact that $c_B$ and $c_M$ induce opposite or same sign changes for the two kinds of power spectra means that combining them will help to break the degeneracy between $c_B$ and $c_M$.
}
\label{fig:cl}
\end{figure}

We show in Fig.~\ref{fig:wk} the impact on the lensing kernel $W^{\kappa}$ by separately varying $c_M$ (orange dashed) and $c_B$ (green dotted) from the fiducial model (blue solid) by $\Delta c = +1.0$ in the case of no tomography. The differences are caused by the different time evolution of $\Sigma$ in the two models, such that the lensing kernel is effectively weighted higher or lower. 

In Fig.~\ref{fig:cl} we show the same effects on the power spectra. Note that the MG effects through the lensing kernel is bigger than that through the modified growth in the matter power spectrum, so we see again positive (negative) shifts for increased $c_B$ ($c_M$) for the lensing-related power spectra $C\kk$ and $C\gk$. On the one hand, $C^{gg}$ is only sensitive to the matter power spectrum, on which the effects of $c_B$ and $c_M$ are much smaller and of the same sign. This would lead to negatively correlated $c_B-c_M$ constraints with $C^{gg}$ and positively correlated constraints with $C\kk$ and $C\gk$. We expect therefore that combining all three probes could break parameter degeneracy in the $c_B-c_M$ plane. Of course, after marginalizing over other cosmological parameters, the degeneracy directions shall become less sharply contrasted, but enough differences remain to yield improved constraints, as we shall see in section~\ref{sec:results}.

\section{Bispectrum Observables}
\label{sec:bis}

\subsection{Definitions}
\label{sec:bis_def}

We follow Ref.~\cite{Takada:2003ef} for the treatment of lensing convergence bispectrum and generalize to the case of cross-bispectra between any three observables 
\beq
A, B, C \in \mathcal{O} \equiv \{\kappa_{(i)}, g_{(i)}\; |\; i = 1\, ..\, n_{\mathrm{zbin}}\}
\eeq
where the fields $\kappa$ and $g$ in different redshift bins are treated as distinct observables.

For the bispectrum, we model the matter density fluctuation up to second-order
\bea
\delta_m(\bm{k}) &=& \delta_{m}^{(1)}(\bm{k}) + \int \frac{d^3 \bm{q}}{(2\pi)^3} \delta_{m}^{(1)}(\bm{q}) 
\delta_{m}^{(1)}(\bm{k} - \bm{q}) 
F_2(\bm{q}, \bm{k} - \bm{q}) \notag\\
&+& \mathcal{O}((\delta_{m}^{(1)})^3).
\eea
and the galaxies as a biased tracer of the matter field up to second order as well
\beq
\delta_g(\bm{x}) = b_1 \delta_m(\bm{x}) + \frac{b_2}{2}\delta_m^2(\bm{x}), 
\eeq
so that in Fourier space
\bea
\delta_g(\bm{k}) &=& b_{1}\delta_{m}^{(1)}(\bm{k}) \notag \\
&+& b_{1} \int \frac{d^3 \bm{q}}{(2\pi)^3} \delta_{m}^{(1)}(\bm{q}) 
\delta_{m}^{(1)}(\bm{k} - \bm{q}) 
F_2(\bm{q}, \bm{k} - \bm{q}) \notag\\
&+& b_{2} \int \frac{d^3 \bm{q}}{(2\pi)^3} \delta_{m}^{(1)}(\bm{q}) 
\delta_{m}^{(1)}(\bm{k} - \bm{q}) 
+\mathcal{O}\left(\delta_{m}^{(3)}\right), \notag\\
\eea
where 
\beq
\label{eq:F2}
F_2(\vk_1, \vk_2)  = \frac{5}{7} + \frac{1}{2} \left( \frac{k_1}{k_2} + \frac{k_2}{k_1} \right) \frac{\vk_1 \cdot \vk_2}{k_1 k_2 } + \frac{2}{7}\frac{(\vk_1 \cdot \vk_2)^2}{k_1 k_2 }
\eeq
is the second-order perturbative kernel in GR.

The full-sky bispectra of the projected quantities $X, Y, Z \in \{\kappa, g\}$ in redshift bins $i,j,k \in \{1, ..., n_{\rm zbin}\}$ respectively is defined as
\begin{eqnarray}
\skaco{X_{l_1m_1(i)}Y_{l_2m_2(j)}
Z_{l_3m_3(k)}}
=\left(
\begin{array}{lll}
l_1&l_2&l_3\\
m_1&m_2&m_3
\end{array}
\right)B_{(ijk)l_1l_2l_3}^{XYZ}, \notag \\
\label{eq:bis_def_full_sky}
\end{eqnarray}
where $\left(
\begin{array}{ccc}
l_1&l_2&l_3\\
m_1&m_2&m_3
\end{array}
\right)$ is the Wigner-3$j$ symbol that describes the coupling between the different modes. We approximate the Wigner-3$j$ symbols by expanding the Stirling approximation to second order (see full expression in Appendix~\ref{sec:wigner}), which is computationally fast and eliminates the accuracy problem for degenerate triangles in the commonly-used first-order expression. 

The full-sky bispectrum is computed using the approximation that relates it to the flat-sky bispectrum, 
\begin{eqnarray}
B_{(ijk)l_1l_2l_3}^{XYZ}&\approx&
\left(
\begin{array}{lll}
l_1&l_2&l_3 \notag \\
0&0&0
\end{array}
\right)
\sqrt{\frac{(2l_1+1)(2l_2+1)(2l_3+1)}{4\pi}} \\
&& B_{(ijk)}^{XYZ}(\bm{l}_1,\bm{l}_2,\bm{l}_3),
\label{eq:rel_flat_to_full}
\end{eqnarray}
where the flat-sky bispectrum is defined as
\beq
\skaco{X_{(i)}(\bm{l}_1)Y_{(j)}(\bm{l}_2)
Z_{(k)}(\bm{l}_3)} =(2\pi)^2
B_{(ijk)}^{XYZ}(\bm{l}_1,\bm{l}_2,\bm{l}_3)\delta^D(\bm{l}_{123}),
\label{eq:bis_def_flat_sky}
\eeq
where $\delta^D(\bm{l}_{123}) = \delta^D(\bm{l}_{1} +  \bm{l}_2 + \bm{l}_3)$.
In the presence of second-order galaxy bias, the flat-sky bispectrum is composed of two pieces
\begin{eqnarray}
B_{(ijk)}^{XYZ}&&(\bm{l}_1,\bm{l}_2,\bm{l}_3)=
\biggl[ \int^{\chi_H}_0\!\!d\chi
W_{(i)}^X \!(\chi)W_{(j)}^Y \!(\chi)W_{(k)}^Z \!(\chi) 
\chi^{-4}\notag \\
&& B_m\!\left(\bm{k}_1,\bm{k}_2,\bm{k}_3;
\chi\right) \biggr] + B_{(ijk),\, b_2}^{XYZ}(\bm{l}_1,\bm{l}_2,\bm{l}_3),
\label{eq:fbispdef}
\end{eqnarray}
where $\bm{k}_i=\bm{l}_i/\chi$ in the Limber approximation.
The first term is a projection of three-dimensional matter bispectrum at tree-level where
\bea
\label{eq:Bm}
B_{m}(\vk_1, \vk_2, \vk_3) = 2P_m(k_1)P_m(k_2) F_2(\vk_1, \vk_2) 
+ 2 \mathrm{\ perm.}\, , \notag \\
\eea
and where $F_2(\vk_1, \vk_2)$ is defined in Eq.~\ref{eq:F2} for GR and shall be modified for MG theories in section~\ref{sec:bis_mg}.

The second term comes from the second-order galaxy bias $b_2$
\beq
B_{(ijk),\, b_2}^{XYZ}(\bm{l}_1,\bm{l}_2,\bm{l}_3) = \int^{\chi_H}_0\!\!d\chi
\chi^{-4} \mathcal{I}^{XYZ}_{(ijk)\,b_2}\!\left(\bm{k}_1,\bm{k}_2,\bm{k}_3;
\chi\right),
\eeq
where
\begin{eqnarray}
\mathcal{I}^{XYZ}_{\,b_2}(\bm{k}_1, \bm{k}_2, \bm{k}_3) 
&=&W^{X}_{(i),b_2} W^{Y}_{(j)} \;\;\, W^{Z}_{(k)}
P(\bm{k}_2) P(\bm{k}_3) \notag \\
&+&  W^{X}_{(i)}\;\;\, W^{Y}_{(j), b_2} W^{Z}_{(k)}
P(\bm{k}_1) P(\bm{k}_3) \notag \\
&+&  W^{X}_{(i)}\;\;\, W^{Y}_{(j)}\;\;\, W^{Z}_{(k),b_2}
P(\bm{k}_1) P(\bm{k}_2). \notag \\
\label{eq:bispectrum_b2}
\end{eqnarray}
The kernels $W^{g}_{(i)}$ and $W^{\kappa}_{(i)}$ involving $b_1$ are given by Eqs.~\ref{eq:Wg_i} and~\ref{eq:Wk_i} respectively, while those involving $b_2$ are given by
\beq
W^{g}_{(i),b_2} = b_{2,(i)} \frac{p_{(i)}^{\rm lens}(z(\chi))}{\bar{n}_{(i)}^{\rm lens}} \frac{dz}{d\chi},
\eeq
and
\beq
\label{eq:Wkappa_b2}
W^{\kappa}_{(i), b_2} = 0.
\eeq
As a result of Eq.~\ref{eq:Wkappa_b2}, all $B^{\kkk}_{(ijk), b_2} = 0$. Furthermore, there is only one non-zero term for $\kappa_{(i)}\kappa_{(j)}g_{(k)}$ and its permutations, e.g.
\begin{eqnarray}
\mathcal{I}^{\kkg}_{(ijk),\,b_2}(\bm{k}_1, \bm{k}_2, \bm{k}_3)=
W_{(i)}^{\kappa} W_{(j)}^{\kappa} W_{(k),\,b_2}^g 
P(\bm{k}_1) P(\bm{k}_2),\notag\\
\label{eq:fbispdef_b2_kkg}
\end{eqnarray}
two terms for $\kappa_{(i)}g_{(j)}g_{(k)}$ and its permutations, e.g.
\begin{eqnarray}
&&\mathcal{I}^{\kgg}_{(ijk),\,b_2}(\bm{k}_1, \bm{k}_2, \bm{k}_3) = W_{(i)}^{\kappa} \times
\notag \\ 
&&\biggl(
W_{(j),\,b_2}^g W_{(k)}^g  
 P(\bm{k}_1)P(\bm{k}_3) 
+
W_{(j)}^g W_{(k),\,b_2}^g  P(\bm{k}_1)P(\bm{k}_2) \biggr) ,\notag\\
\label{eq:fbispdef_b2_kgg}
\end{eqnarray}
and three for $g_{(i)}g_{(j)}g_{(k)}$. 

Modelled as such, we have chosen to compute the lowest-order (non loops) terms of the power spectra and the bispectra -- $P_{11}$ and the tree-level bispectrum respectively. Note that $b_2$ did not appear in section~\ref{sec:ps_def} because it does not enter $P_{11}$. We have also chosen to ignore $b_{s2}$ for simplicity.

\begin{widetext}

The covariance between two general bispectra is given by the Wick's theorem as
\begin{eqnarray}
f_{\rm sky}{\rm Cov}\left[B_{l_1l_2l_3(ijk)}^{XYZ},
B_{l_1^\prime l_2^\prime l_3^\prime(abc)}^{X'Y'Z'}\right] 
&\approx& 
\tilde{C}_{(ia)}^{XX'}(l_1)\delta_{l_1l_1^{\prime}}
\left[
\tilde{C}_{(jb)}^{YY'}(l_2)
\tilde{C}_{(kc)}^{ZZ'}(l_3)\delta_{l_2l_2^{\prime}}\delta_{l_3l_3^{\prime}}
+\tilde{C}_{(jc)}^{YZ'}(l_2)
\tilde{C}_{(kb)}^{ZY'}(l_3)\delta_{l_2l_3^{\prime}}\delta_{l_3l_2^{\prime}}
\right]\nonumber\\
&+&
\tilde{C}_{(ib)}^{XY'}(l_1)\delta_{l_1l_2^{\prime}}
\left[
\tilde{C}_{(ja)}^{YX'}(l_2)
\tilde{C}_{(kc)}^{ZZ'}(l_3)\delta_{l_2l_1^{\prime}}\delta_{l_3l_3^{\prime}}
+\tilde{C}_{(jc)}^{YZ'}(l_2)
\tilde{C}_{(ka)}^{ZX'}(l_3)\delta_{l_2l_3^{\prime}}\delta_{l_3l_1^{\prime}}
\right]\nonumber\\
&+&
\tilde{C}_{(ic)}^{XZ'}(l_1)\delta_{l_1l_3^{\prime}}
\left[
\tilde{C}_{(ja)}^{YX'}(l_2)
\tilde{C}_{(kb)}^{ZY'}(l_3)\delta_{l_2l_1^{\prime}}\delta_{l_3l_2^{\prime}}
+\tilde{C}_{(jb)}^{YY'}(l_2)
\tilde{C}_{(ka)}^{ZX'}(l_3)\delta_{l_2l_2^{\prime}}\delta_{l_3l_1^{\prime}}
\right],
\label{eq:covbisp}
\end{eqnarray}
where we have ignored non-Gaussian terms from connected 3-, 4-, and 6-point functions following Ref.~\cite{Takada:2003ef} which verified that these terms are expected to be small over the angular range considered here for the lensing convergence bispectrum, which is the one that dominates our results as we shall see in section~\ref{sec:results}. 
\end{widetext}

The Kronecker delta functions in Eq.~\ref{eq:covbisp} enforce that the different triangles are uncorrelated. They also enforce that only one of the six terms is non-zero for a general triangle $l_1 \neq l_2 \neq l_3$, while two terms are present for isoceles triangles and all six for the equilateral triangles. Note that unlike for the auto-bispectrum, when considering the cross-bispectrum of different observables (e.g. $B^{\kappa gg}_{(123)}$), these two or six terms are not necessarily equal to each other anymore, so the full expression above shall be used, rather than the auto-bispectrum version:
\bea
&&f_{\rm sky} {\rm Cov}\left[B_{l_1l_2l_3(ijk)}^{XXX},
B_{l_1^\prime l_2^\prime l_3^\prime(abc)}^{XXX}\right] \notag \\
&\approx& \Delta(l_1,l_2,l_3) 
C_{(ia)}^{XX}(l_1)C_{(jb)}^{XX}(l_2)
C_{(kc)}^{XX}(l_3)\delta_{l_1l_1^\prime}
\delta_{l_2l_2^\prime}\delta_{l_3l_3^\prime}, \notag \\
\label{eqn:covbisp2}
\eea
where $\Delta(l_1, l_2, l_3) = 1$, 2, or 6 for general, isoceles and equilateral triangles.

In practice, we do not use all six terms but only keep the first of them for calculating the Fisher matrix. This is equivalent to treating all triangles as a general triangle $l_1 \neq l_2 \neq l_3$ even if they were actually equilateral or isoceles. The motivation behind this is that as we bin in $l_1$ and $l_2$ in the Fisher section, we expect that most of the triangles in a given $l-$bin represented by the bin-center are not exactly equilateral or isoceles even if the triangle at the bin center happens to be one. 

In principle, there are also additional non-Gaussian in-survey and super-sample terms in the covariance between the bispectra as in the power spectrum case. In Ref.~\cite{Rizzato:2018whp}, it was shown that including these terms lead to at most a factor of $2-3$ degradation in the signal-to-noise for combined lensing power spectrum and bispectrum for the range $l_{\rm max} = 1000-3000$, which further reduces to a factor $\lesssim2$ when tomography is used. A similar argument as the one made for the power spectrum constraints in section~\ref{sec:ps_def} applies here as well: the projected 1D error on parameters would likely not exceed 15\% when the higher-dimensional volume change by a factor of $2-3$ for a Fisher analysis with 8 or more parameters.

\subsection{Bispectrum in Modified Gravity Theories}
\label{sec:bis_mg}

As described in section~\ref{sec:ps_def}, the effects of MG on the power spectrum observables mainly come through a modified growth of perturbations and gravitational slip which alters the matter power spectrum and the lensing kernel respectively. These effects can be described by phenomenological parameters $\mu$ and $\Sigma$ which are related to the $\alpha_i$ parameters in Horndeski theories. For the bispectrum, there is an additional effect through the second-order perturbative kernel $F_2$ parameterized by $\lambda$~\cite{Yamauchi:2017ibz}:
\bea
\label{eq:F2_MG}
F_2(\bk_1, \bk_2) &=& 1 + \frac{1}{2} \left( \frac{k_1}{k_2} + \frac{k_2}{k_1} \right) \frac{\vk_1 \cdot \vk_2}{k_1 k_2 }  \notag \\
&-& \frac{2}{7} \lambda(a) \left( 1- \frac{(\bk_1 \cdot \bk_2)^2}{k_1^2 k_2^2}\right),
\eea 
where $\lambda(a)$ obeys a second-order differential equation~\cite{Takushima:2015iha, Takushima:2013foa, Yamauchi:2017ibz} 
\bea
\label{eq:lambda eq}
&&\frac{\dd^2\lambda}{\dd\ln a^2}+\left( 2+\frac{\dd\ln H}{\dd\ln a}+4f\right)\frac{\dd\lambda}{\dd\ln a}
		+\left( 2f^2+\kappa_\Phi\right)\lambda \notag \\
		&=&\frac{7}{2}\left( f^2-\tau_\Phi\right),
	\,
\eea
and $\lambda = 1$ in GR. Here $H$ is the Hubble parameter, $f$ is the linear growth rate whose evolution is also modified~\cite{Yamauchi:2017ibz}
\beq
\frac{\mathrm{d}f}{\mathrm{d\,ln\,}a } 
+ \left( 2 + \frac{\mathrm{d\,ln\,}H }{\mathrm{d\,ln\,} a } \right)\,f \, + \, f^2 - \kappa_{\Phi} = 0,
\eeq

and $\kappa_{\Phi}$ and $\tau_{\Phi}$ encode the MG modifications to the usual equation describing gravitational evolution,

\bea
\label{eq:Phi assumption}
	-\frac{k^2}{a^2H^2}\Phi (a,{\bm k})&=&\kappa_\Phi (a,k)\, \delta (a,{\bm k})\notag \\
		&+&\int\frac{\dd^3{\bm k}_1\dd^3{\bm k}_2}{(2\pi )^3}\, 
		\delta^3_{\rm D}({\bm k}_1+{\bm k}_2-{\bm k})\, \tau_\Phi (a) \notag \\
		&&\times \left(
	1-\frac{({\bm k}_1\cdot{\bm k}_2)^2}{k_1^2k_2^2} \right)
		 \delta (a,{\bm k}_1)\,\delta (a,{\bm k}_2)	\notag \\
		&+& \mathcal{O}(\delta^3)
	\,.
\eea

Instead of solving the differential equation for $\lambda(a)$, we follow Ref.~\cite{Yamauchi:2017ibz} to use a phenomenological parametrization 
\beq
\lambda(a) = \tilde{\Omega}_m^{\xi}(a),
\eeq
where $\tilde{\Omega}_m(a)$ is the evolution of the matter density parameter and $\xi$ is a fixed exponent. To leading order, $\xi$ takes the following form
\bea
	\xi
		=\frac{-3+6\gamma+2\kappa_\Phi^{(1)}+7\tau_\Phi^{(1)}}{(7-6w^{(0)}+2c_{\rm M})(1-3w^{(0)}+c_{\rm M})}
	\,,\label{eq:Gamma estimation}
\eea
where $\tilde{\gamma}$ is the gravitational growth index 
$f \approx \widetilde\Omega_m^{\tilde{\gamma}}(a)$ and $\tilde{\gamma} \approx 0.55$ in GR. 
The expressions for $\tilde{\gamma}$ in MG as well as for the lowest order expansion coefficients $\kappa_{\Phi}^{(1)}$, $\tau_{\Phi}^{(1)}$, $w^{(0)}$ as a function of $c_i$ are given in Appendix~\ref{sec:lambda} which are reproduced from Ref.~\cite{Yamauchi:2017ibz}.

\subsection{Comments on the limitations of the adopted bispectrum modeling}

There are a few limitations to the prescription used above to model MG effects on the bispectrum.

First, the prescription of modifying $F_2$ with $\lambda(a)$ is only valid for models in which the growth is a function of time alone (e.g. not valid for models like $f(R)$ where the growth is also scale-dependent). In Ref.~\cite{Alonso:2016suf}, the effects of screening on the power spectrum was modelled phenomenologically by introducing scale-dependent $\alpha$'s with a cut-off scale: $\alpha_i \rightarrow \alpha_i S(k/k_v)$ where $S(k/k_v) = \mathrm{exp}\left(-\frac{1}{2}(k/k_v)^2\right)$ where $\alpha$'s return to their GR values for scales smaller than the cut-off scale $\pi/k_v$. This would give the scale-dependence that renders invalid the $F_2$ prescription adopted here for our bispectrum modeling.

Because any realistic MG models must pass the solar system tests with possibly a screening mechanism that returns the theory to GR on small-scales, one might worry that not accounting for the screening would overestimate the amount of signal there is in reality on the small-scales. It was however shown in Ref.~\cite{Alonso:2016suf} that introducing a screening scale as described above actually yields better constraints on parameters, as the existence of a new scale ends up contributing to break the degeneracies with other parameters. So although neglecting the screening effects here means not modeling the small scales accurately enough, it would actually lead to a more conservative, rather than optimistic forecast. 

Second, the bispectrum modeling used here only includes the tree-level contribution, which is valid up to roughly $k\sim 0.1\, h^{-1}\mathrm{Mpc}$. For GR, a well-tested extension into the nonlinear regime exists where coefficients in front of the various terms in $F_2$ are added and fitted to GR simulations~\cite{Scoccimarro:2000ee}. A similar extension into the nonlinear regime for MG is still being tested. 

In Ref.~\cite{Bose:2019wuz}, the authors combined the $\lambda(a)$ prescription together with the GR fitting formula to model the bispectrum in the nonlinear regime as
\bea
B^{\rm fit}(\bm k_1, \bm k_2,\bm k_3;a) &=& 2 P_m^{\rm NL}(k_1;a) P_m^{\rm NL}(k_2;a) F^{\rm fit}_2(\bm k_1,\bm k_2;a) \notag\\
&+& \mathrm {2 \; perm.}\,,  
\label{eq:bis_nl} 
\eea
where, compared to Eq.~\ref{eq:Bm}, the linear matter power spectrum has now been replaced by the non-linear power spectrum $P_m^{\rm NL}$ in MG, and where the $F_2$ kernel now includes non-linear effects through the coefficients $\bar{a}, \bar{b}$, and $\bar{c}$ which are fitted on GR simulations:
\bea
\label{eq:F2_NL}
F^{\rm fit}_2(\bm k_1,\bm k_2;a) &=& \left( 1 - \frac{2}{7}  \lambda(a) \right)
\, \bar{a}(k_1,a)\, \bar{a}(k_2,a)
\notag \\ 
&+& \frac{1}{2} \left( \frac{k_1}{k_2} 
+ \frac{k_2}{k_1} \right) \frac{\vk_1 \cdot \vk_2}{k_1 k_2 } \,\bar{b}(k_1,a)\,\bar{b}(k_2,a) \notag \\
&+& \frac{2}{7}\, \lambda(a)\, \frac{(\bk_1 \cdot \bk_2)^2}{k_1^2 k_2^2} \, \bar{c}(k_1,a)\, \bar{c}(k_2,a). \notag \\
\eea
The validity of this formula is tested against simulations in Ref.~\cite{Bose:2019wuz} for the $f(R)$ and the DGP models in the equilateral triangle configurations. More validation work is to be done for other MG models as well as for general triangle configurations. While this work is in progress, we restrict ourselves to the modeling at the tree-level which becomes less valid in the non-linear regime. We will control the degree to which this affect our results by varying the angular scale cuts $l_{\rm max}$ of our Fisher results in section~\ref{sec:results}, and note that we expect better constraints once the non-linear regime can be properly modelled.

\begin{figure*}[ht]
\begin{minipage}{0.93\linewidth}
\centering
\includegraphics[width=0.93\linewidth]{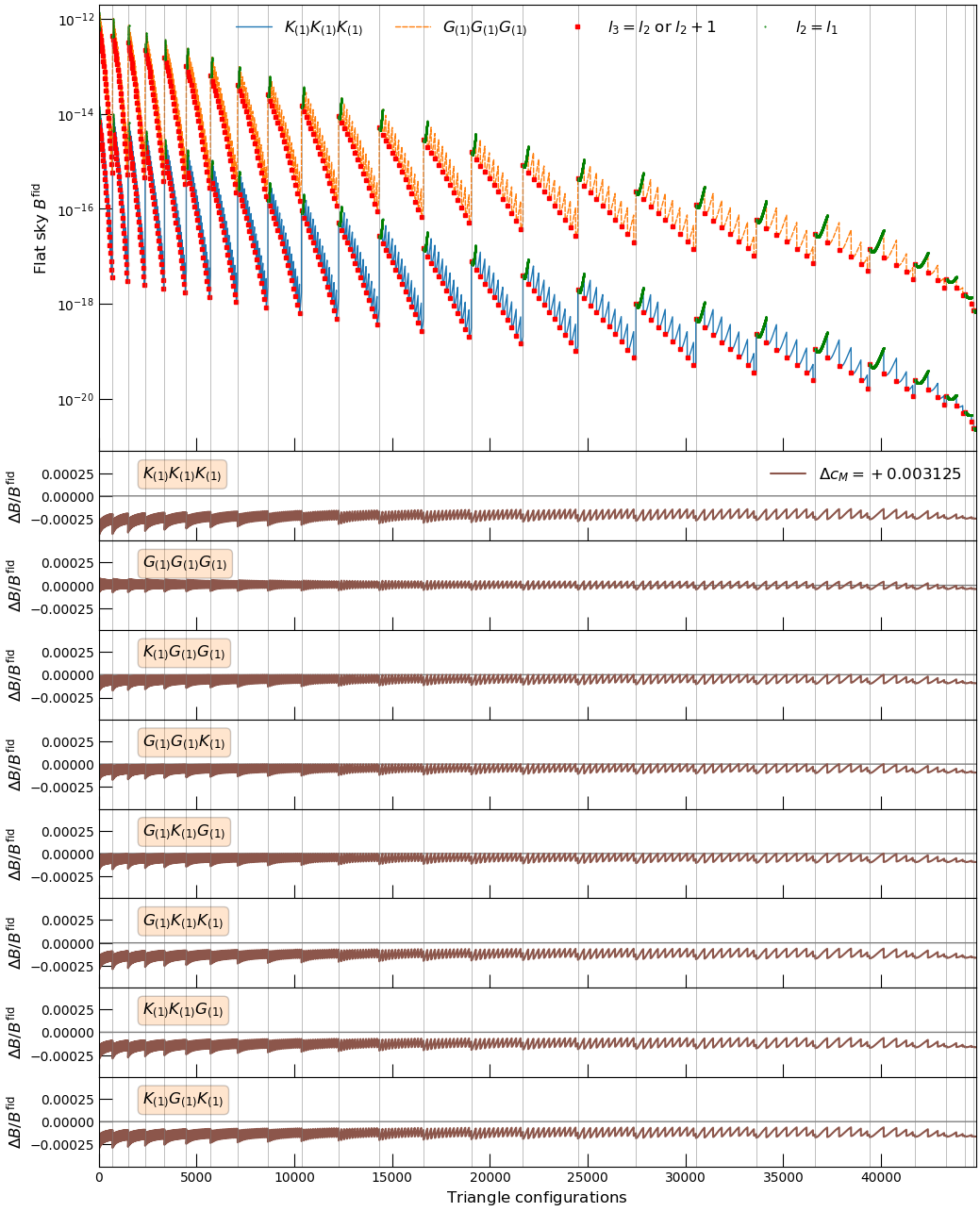}
\caption{
\textit{Top panel}: Flat sky bispectrum signal in the fiducial model vs. triangle configurations for two out of eight bispectra combinations for $n_{\rm zbin} = 1$:  $\kappa_{(1)}\kappa_{(1)}\kappa_{(1)}$ (blue solid) and $g_{(1)}g_{(1)}g_{(1)}$ (orange dashed); all other bispectra (not shown) have similar shapes but with amplitudes somewhere between these two. We order the triangle configurations first by increasing $l_1$ ($l_1$ steps up at grey vertical lines), then $l_2$ (which steps up or down at red squares) and then $l_3$. The red squares also mark the isoceles triangles $l_3 = l_2$ or near-isoceles triangles $l_3 = l_2 + 1$ (when the bispectra with $l_3 = l_2$ is zero due to vanishing Wigner-3j symbols for $l_1+l_2+l_3 =$ odd); the green dots mark the $l_2 = l_1$ isoceles triangles and for a given $l_1$, they form a line along which $l_3$ is increased. \\
\textit{Lower panels}: Fractional deviations from the fiducial bispectrum signal for all eight bispectra in the $n_{\rm zbin} = 1$ case, when $c_M$ is varied by $\Delta c_M = 0.003125$ (as used in the derivative computation).}
\label{fig:bis_signal_nzbin1_cM}
\end{minipage}
\end{figure*}

\begin{figure*}[ht]
\begin{minipage}{0.93\linewidth}
\centering
\includegraphics[width=0.93\linewidth]{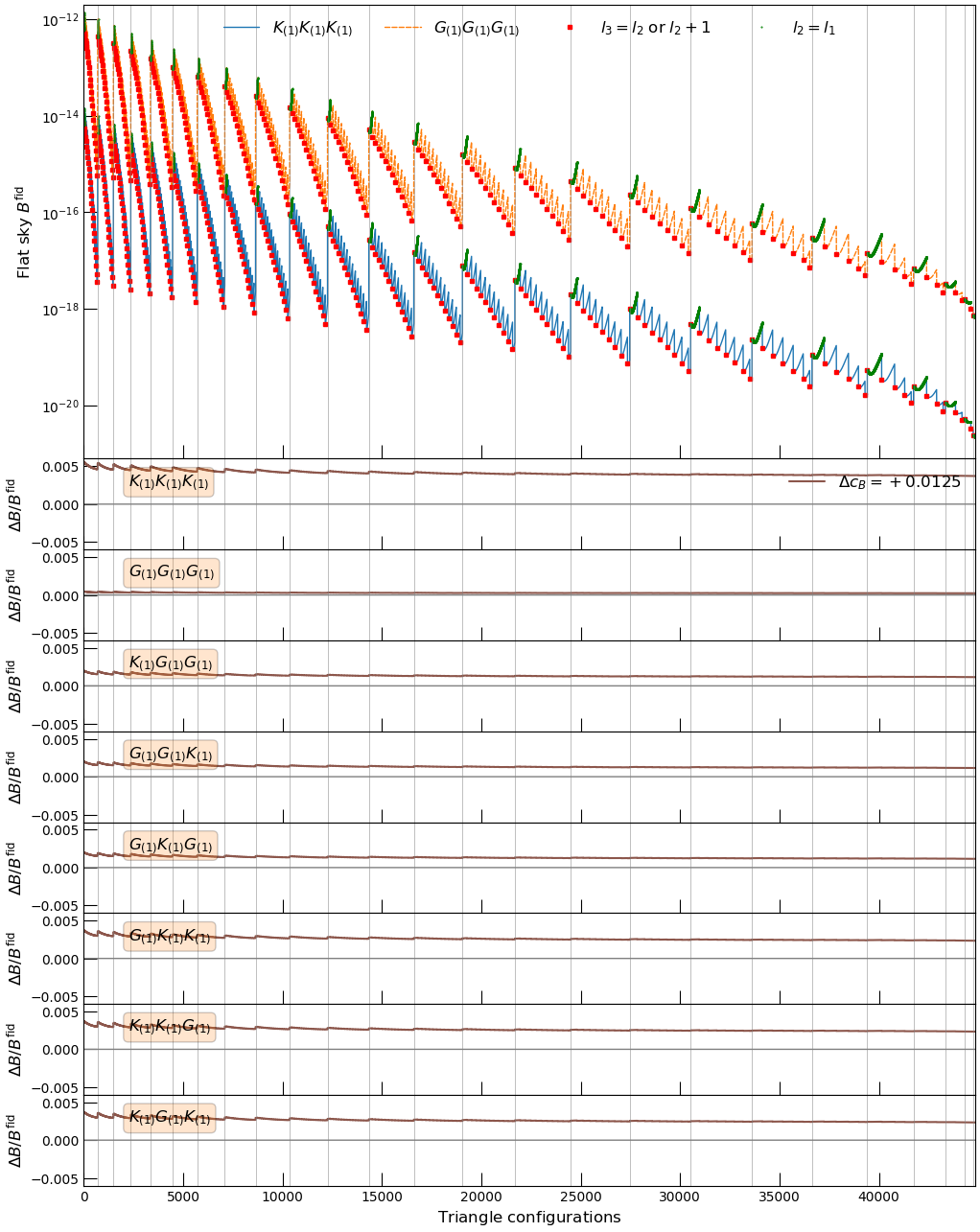}
\caption{Same as Fig.~\ref{fig:bis_signal_nzbin1_cM}, but for $c_B$ varying by $\Delta c_B$ = 0.0125. }
\label{fig:bis_signal_nzbin1_cB}
\end{minipage}
\end{figure*}

\begin{figure*}[ht]
\begin{minipage}{0.93\linewidth}
\centering
\includegraphics[width=0.93\linewidth]{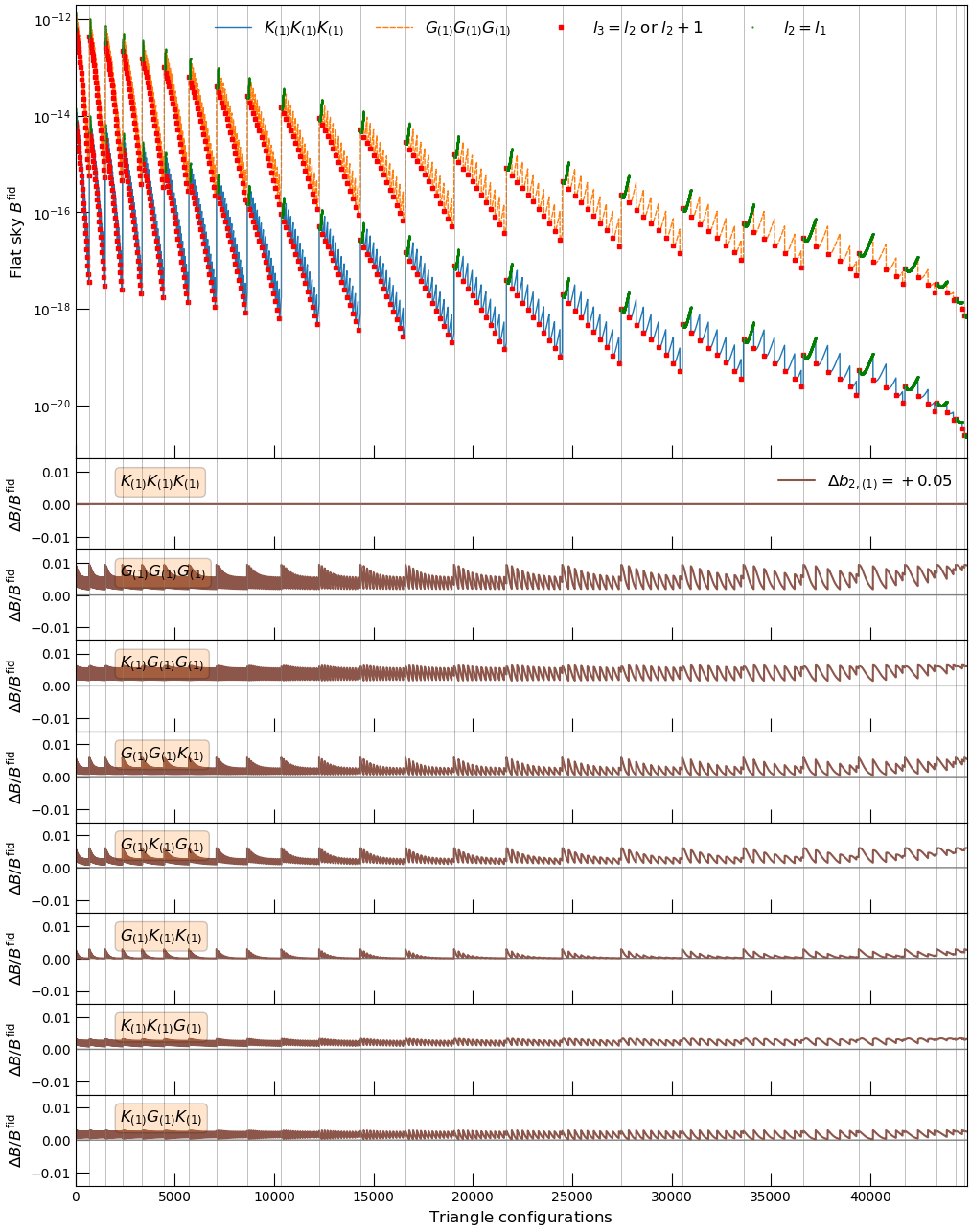}
\caption{Same as Fig.~\ref{fig:bis_signal_nzbin1_cM}, but for $b_{2,(1)}$ varying by $\Delta b_{2,(1)} = 0.05$. Note that while the fractional deviations are the same for cyclically permuted bispectra (panels 4-6 and 7-9) for $c_M$ and $c_B$, it is not so when $b_{2,(1)}$ is varied.}
\label{fig:bis_signal_nzbin1_b2z1}
\end{minipage}
\end{figure*}

\section{Forecast Setup}
\label{sec:fisher}

We now describe the Fisher matrix formalism used to obtain the results later presented in section~\ref{sec:results}. We first set up the set of observables to be used in section~\ref{sec:unique_observables} and describe the Fisher matrix formulae in section~\ref{sec:fisher_matrix}.

\subsection{The complete set of unique bispectra}
\label{sec:unique_observables}

One needs to be careful when finding the unique and complete set of cross-bispectrum observables when using tomography. The $\kappa$ and $g$ fields in each redshift bin is now counted as a unique observable. So for $n_{\mathrm{zbin}} = 3$ redshift bins and 2 fields, we really have 6 different observables 
$\mathcal{O} = \{ \kappa_{(i)}, g_{(i)}\, | \,i = 1\, ..\, n_{\mathrm{zbin}} \}$.
The bispectra between them can then be separated into three categories: 

\begin{enumerate}
\item{Pure auto-bispectra, such as $\kappa_{(1)}\kappa_{(1)}\kappa_{(1)}$ or $g_{(2)}g_{(2)}g_{(2)}$;}
\item{Cross-bispectra where two of the observables are the same, such as $\kappa_{(1)}g_{(2)}g_{(2)}$ or $\kappa_{(3)}\kappa_{(3)}g_{(1)}$;}
\item{Cross-bispectra between three completely different observables, such as $\kappa_{(1)}\kappa_{(2)}g_{(3)}$.}

\end{enumerate}

Note that if we counted only unique triangles $\{(l_1, l_2, l_3)\, |\, l_1 \leq l_2 \leq l_3 \}$ for each combination $ABC$ where $A, B, C \in \mathcal{O}$, then in case 1 ($A = B = C$), any of the six possible permutations of $ABC$ would be redundant. However, in case 3 (where all three observables $A, B$ and $C$ are distinct), then all six permutations of $ABC$ are unique. Finally, we have case 2 which are the intermediate cases (either $A = B$, $A = C$ or $B = C$), where the three \textit{cyclic} permutations of $ABC$ form a unique set. To account for all of this, we adopt only the unique permutations of $ABC$ for each case as described above. It would be equivalent to permute $(l_1, l_2, l_3)$ instead of $ABC$, but because this would result in different sets of triangles to be looped over for each kind of bispectrum, we find it easier in practice to not do so. We will implement the loop over multipoles as an
``outer loop".

In general, there are a total of $n_{\rm obs}^3$ distinct bispectra; for $n_{\rm zbin} = 3$ this would be 216. We can however reduce this number in our case by noticing that $g_{(i)}g_{(j)}g_{(k)}$ is only nonzero if we are considering the same redshift bin $i = j = k$. Because the lensing kernels are nonzero over the redshift range up to the bin considered, a further reduction can be done by keeping only the bispectra in which $\kappa$ are not from a redshift bin lower than the lowest bin for any $g$. This reduces the total number of bispectra to model to 90 for $n_{\rm zbin} = 3$ and 34 for $n_{\rm zbin} = 2$. 

We note also that often times in the literature, a redefinition of the bispectrum in case 2 is used, when the bispectrum it is invariant under cylic permutations of $ABC$. For example, for
\beq
B^{\kkg}_{(123)}(l_1, l_2, l_3) = B^{g\kappa\kappa}_{(312)}(l_1, l_2, l_3) = B^{\kappa g \kappa}_{(231)}(l_1, l_2, l_3),
\label{eq:cyclic_perm}
\eeq
one would possibly redefine $B^{\kkg}_{(123)}(l_1, l_2, l_3)$ to mean just the sum of all three bispectra
\beq
B^{\kkg}_{(123)}(l_1, l_2, l_3) \rightarrow 3B^{\kkg}_{(123)}(l_1, l_2, l_3),
\eeq
and deal with less bispectrum observables. In this work, however, the invariance is broken by the presence of the second-order bias $b_2$ terms\footnote{Note that the sum of those three terms in Eq.~\ref{eq:cyclic_perm} can still be expressed as invariant under cyclic permutations as long as $b_2$ and $b_1$ do not evolve with redshift. Here this form is explicitly broken as we model $b_2$ and $b_1$ to change between tomographic bins.}. 
But even if we were to not include those $b_2$ terms, because the covariance with other bispectra is not invariant under cyclic permutations (need to match $l_1$ with $l_1$, etc), we would still need to spell out the individual bispectrum in the definition. For these reasons, we adopt the conventions described above.

In Figs.~\ref{fig:bis_signal_nzbin1_cM}, ~\ref{fig:bis_signal_nzbin1_cB} and~\ref{fig:bis_signal_nzbin1_b2z1}, we show the impact of varying the parameters $c_M$, $c_B$ and $b_2$ on all eight bispectra found in the case of $n_{\rm zbin} = 1$. In the top panel, we show the flat sky bispectrum signal in the fiducial model as a function of triangle configurations for  $\kappa_{(1)}\kappa_{(1)}\kappa_{(1)}$ (blue solid) and $g_{(1)}g_{(1)}g_{(1)}$ (orange dashed). We do not show the other bispectra as they all have similar shapes with amplitudes between the two shown curves.

The triangle configurations are ordered by increasing $l_1$, $l_2$, then $l_3$.  The grey vertical lines denote where $l_1$ steps up; the red squares where $l_2$ is steps up or down, they also correspond to the isoceles triangles with $l_3 = l_2$ or near-isoceles triangles with $l_3 = l_2 + 1$ (for when the bispectra with $l_3 = l_2$ is zero due to Wigner-3j symbols vanishing for $l_1+l_2+l_3 =$ odd); finally the green dots denote $l_2 = l_1$ isoceles triangles.

We show in the lower panels the fractional deviation from the fiducial bispectrum signal. For $c_M$ and $c_B$, all three cyclic permutations of case 2 $ABC$ (panels 4-6 and 7-9) have the same fractional deviations; this is not true for the $b_{2,(1)}$ parameter as mentioned above.

\begin{widetext}

\subsection{The Fisher matrix formalism}
\label{sec:fisher_matrix}

Given a set of bispectra $\mathcal{M}$, the Fisher matrix from combining all of them is given by
\begin{eqnarray}
\label{eq:fbis}
F_{\alpha\beta}^{\rm B} =
\sum_{l_{\rm min}\le 
l_1\le l_2\le l_3\le l_{\rm max}} \ 
\sum_{M, M' \in \mathcal{M}}
\frac{\partial 
B_{l_1l_2l_3}^{M}}{\partial p_\alpha}
\left[{\rm Cov}
[B_{l_1l_2l_3}^{M},
B_{l_1l_2l_3}^{M'}
]\right]^{-1}
\frac{\partial B_{l_1l_2l_3}^{M'}}{\partial p_\beta}. \notag\\
\end{eqnarray}
where we considered the parameters $p_\alpha$ and $p_\beta$ and ${\rm Cov}[X, Y]$ denote the covariance matrix defined in Eq.~\ref{eq:covbisp}. In this work, we will obtain results for the entire set of unique and non-zero bispectra described in the previous section $\mathcal{M} = \mathcal{M^{\rm tot}}$, as well as for various subsets of $\mathcal{M}^{\rm tot}$ such as those containing only $\kappa$, only $g$, or only auto-bispectra. Note that because of the use of tomography, an auto-bispectrum is no longer anything of the form $\kkk$ or $ggg$, but rather $\kappa_{(i)}\kappa_{(i)}\kappa_{(i)}$ or $g_{(i)}g_{(i)}g_{(i)}$ with fields belonging to the same redshift bin.

For our fiducial results, we will take $n_{\rm zbin} = 3$ for which the covariance matrix is a $90\times90$ matrix for each triplet $(l_1, l_2, l_3)$. Recall that the bispectrum covariance is diagonal in triangle configuration space, so we only need to sum over pairs of the same triangle configuration $(l_1, l_2, l_3)$ and make sure to count each configuration only once by imposing $l_1\le l_2\le l_3$. 

Since it is intractable to compute contributions from every $l$, we follow Ref.~\cite{Takada:2003ef} to bin $l_1$ and $l_2$
\begin{eqnarray}
F^{\rm B}_{\alpha\beta}=
\sum_{l_{\rm min}\le 
\bar{l}_1\le \bar{l}_2\le l_3\le l_{\rm max}}\
\Delta l_1\Delta l_2
\sum_{M, M' \in \mathcal{M}}
\frac{\partial B^{M}_{\bar{l}_1 \bar{l}_2 l_3}}{\partial p_\alpha}
\left[{\rm Cov}[B^{M}_{\bar{l}_1 \bar{l}_2 l_3},
B^{M'}_{\bar{l}_1 \bar{l}_2  l_3}
]\right]^{-1}
\frac{\partial B^{M'}_{\bar{l}_1 \bar{l}_2  l_3}}{\partial p_\beta},
\label{eq:fbis_bin}
\end{eqnarray}
where $\bar{l}_1, \bar{l}_2$ denote the logarithmic center of the logarithmic bins in  $l_1$ and $l_2$. As pointed out by Ref.~\cite{Takada:2003ef}, since the Wigner 3-$j$ symbol is only non-zero for $\bar{l}_1+ \bar{l}_2 + l_3={\rm even}$ and vanishing for $\bar{l}_1+ \bar{l}_2 + l_3={\rm odd}$, and the nonzero values themselves change rapidly in sign when varying $l_3$ with fixed $l_1$, $l_2$, one should not bin over $l_3$ for accurate results.

Note that for the purpose of calculating the Wigner-3$j$ symbols, $\bar{l}_1, \bar{l}_2$ are actually the nearest integer to the log bin-centers, which is a good enough approximation if the bins are large enough to cover multiple integers.
For all our results, we use 26 logarithmic $l$-bins between $l_{\rm min}=50$ and $l_{\rm max}=3000$. 
This is reasonable as the bispectra considered here are smooth enough between the bins chosen.  

For the power spectrum Fisher matrix, we use
\begin{eqnarray}
\label{eq:fps_bin}
F^{\rm PS}_{\alpha\beta}=
\sum_{\bar{l}}
\Delta l \ 
\sum_{N, N' \in \mathcal{N}}
\frac{\partial C^{N}_{\bar{l}}}{\partial p_\alpha}
\left[{\rm Cov}[C^{N}_{\bar{l}},
C^{N'}_{\bar{l}}
]\right]^{-1}
\frac{\partial C^{N'}_{\bar{l}}}{\partial p_\beta},
\end{eqnarray}
where $\mathcal{N} = \{\kappa_{(i)}\kappa_{(j)},\, g_{(i)}\kappa_{(j)} ,\, g_{(i)}g_{(j)}\; |\; i,j = 1,2,...,n_{\rm zbin} \; \mathrm{and}\; i\leq j \}$ is a set of unique auto and cross power-spectra. The covariance matrix at a given $\bar{l}$ is a $|\mathcal{N}| \times |\mathcal{N}|$ matrix given by  Eq.~\ref{eq:covps} where $\mathcal{N} = 3\,n_{\rm zbin} (n_{\rm bin} + 1)/2 $. 

\end{widetext}

To compute the total Fisher information combining the power spectra and bispectra, we simply add the two Fisher matrices and ignore correlations between them
\begin{equation}
F_{\alpha\beta}^{\rm tot}\approx F_{\alpha\beta}^{\rm PS}+F_{\alpha\beta}^{\rm B}. 
\end{equation}
In principle, there are additional correlations arising from the 5-point function between the observables, which could degrade the total constraints. We leave its consideration for future work and focus on our aim of estimating the relevance of cross-bispectra for this work. 
 
Under the approximation that the likelihood function is a multivariate Gaussian, the inverse of the Fisher matrix gives us the covariance between any two measured parameters $\alpha$ and $\beta$
\bea
\mathrm{Cov}[\alpha, \beta] = 
\left[
\begin{array}{cc}
(\bm{F}^{-1})_{\alpha\alpha}&(\bm{F}^{-1})_{\alpha\beta}\\
(\bm{F}^{-1})_{\alpha\beta}&(\bm{F}^{-1})_{\beta\beta}
\end{array}
\right].
\eea
We use this covariance matrix to plot the 2D contours on parameter constraints in section~\ref{sec:results}. Moreover, the 1D marginalized constraints on a parameter $\alpha$ will be given by 
\beq
\sigma_{\alpha} = \left[(\bm{F}^{-1})_{\alpha\alpha}\right]^{1/2}.
\eeq

Throughout this work, we use a $\Lambda$CDM model consistent with the final Planck 2018 results (baseline model 2.5): primordial spectrum amplitude and tilt $A_s = 2.1\times10^{-9}$ and $n_s = 0.966$, Hubble constant $h_0 = 0.673$, matter density $\Omega_m = 0.316 $, and baryon density $\Omega_b = 0.0494$. The derivatives are computed around a modified gravity fiducial model very close to GR as mentioned before: $c_K = 0.1$, $c_B = 0.05$, $c_M = 0.05$, $c_T = 0.05$. This is chosen so that we can avoid numerical singularities at $c_i = 0$'.

The derivatives are calculated using a two-sided finite difference by varying the following set of parameters one at a time from their fiducial values ($c_T$ and $c_K$ are always fixed): 
$\{c_M, c_B, \Omega_m, b_{j, (i)}, A_s, \Omega_b, h_0, n_s \;|\; i= 1, ..., n_{\rm zbin}\}$ where we have only linear galaxy biases $j = 1$ for the power spectra, and both linear and second-order biases $j = 1, 2$ for the bispectra.
The fiducial values for the linear galaxy biases are fixed at $b_{1, (i)} = 1$ for all redshift bins, while the fiducial second-order biases are computed using a fitting formula $b_2(b_1)$ derived from GR simulations (Eq. 5.2 of Ref.~\cite{Lazeyras:2015lgp}) by evaluating it at the fiducial value $b_1 = 1$; however we do not vary $b_2$ with this formula when we vary $b_1$ for the derivative computation so to treat them as separately measured parameters.

 Now most of the parameters are varied with a step size $5\%$ of their fiducial value, except for $c_M$, $c_B$ and $\Omega_m$ where we used $\Delta c_M = 0.003125$, $\Delta c_B = 0.0125$ and $\Delta \Omega_m = 0.004$ to guarantee convergence of the derivatives. For the four parameters at the end of list, we imposed priors consistent with the Planck 2018 constraints: they are $\sigma_{A_s} = 3.1 \times 10^{-11}$, $\sigma_{\Omega_b} = 3.3 \times 10^{-4}$, $\sigma_{h_0} = 0.006$ and $\sigma_{n_s} = 0.0044$. These priors are included by adding a diagonal matrix $(F_{\rm prior})_{\alpha\beta} = \delta_{\alpha \beta} \sigma_{\alpha}^{-2}$.\\

\section{Results}
\label{sec:results}

We will now show the results of the Fisher forecasts.  We will first study the constraints from the bispectra alone in section~\ref{sec:bispectrum_results}, 
by breaking down the results from all bispectra into those from smaller subsets, showing how using all cross-bispectra between $\kappa$ and $g$ in various redshift bins help to improve parameter constraints. 
Then we will look at the total results from combining the power spectra and bispectra in section~\ref{sec:total_results}, as well as the dependence on some forecast parameters: the number of redshift bins and $l_{\rm max}$. Unless otherwise mentioned, we report our results at a fiducial choice of $l_{\rm max}~=~1000$ and $n_{\rm zbin} = 3$.

\subsection{Bispectrum results}
\label{sec:bispectrum_results}
\begin{figure}[t]
\includegraphics[width=0.5\textwidth]{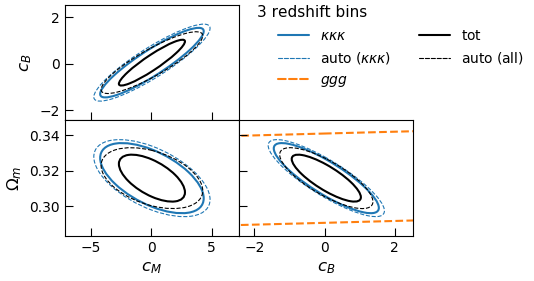}
\caption{Marginalized 2D parameter constraints shown at 68$\%$ confidence level from all bispectra and chosen subsets. Results are shown for our fiducial choice of $l_{\rm max} = 1000$ and $n_{\rm zbin} = 3$, and the 2000 deg$^2$ area of overlap between the notional Roman Space Telescope survey and LSST. The $\kkk$ constraint (blue solid) does much better than the $ggg$ constraint (orange dashed) which is too big to show for some of the panels. Combining them (black solid) results in a significant improvement, a factor of (8, 52, 3) for the constraints on ($c_M$, $c_B$, $\Omega_m$) compared to $ggg$ alone and (1.6, 1.5, 1.5) for $\kkk$ alone. Furthermore, while $\kkk$ result is dominated by the $\kappa$ auto-bispectra, the total bispectrum result is not dominated by the auto-bispectra alone (black dashed), indicating that including cross-bispectra between $\kappa(i)$ and $g(j)$ serves to improve constraints, quantitatively, by a factor of (1.3, 1.1, 1.3) on $c_M$, $c_B$ and $\Omega_m$ constraints respectively.}
\label{fig:breakdown}
\end{figure}

The marginalized 2D parameter constraints for $c_M$, $c_B$ and $\Omega_m$ at 68$\%$ confidence level are shown in Fig.~\ref{fig:breakdown} for the set of all unique and nonzero bispectra as well as for a few informative subsets.

Restricting $\mathcal{M}$ in Eq.~\ref{eq:fbis_bin} to $\mathcal{M} = \{ M \in \mathcal{M}^{\rm tot}\; |\; M = \kappa_{(i)}\kappa_{(j)}\kappa_{(k)} \}$ give the $\kkk$ only results, and similarly for $ggg$ alone. They correspond to the solid blue line and the dashed orange line respectively. The $ggg$ only contours are too big to show in some of the panels -- about $\sim5$ times bigger in the $c_M$ direction and about $\sim35$ times bigger for $c_B$. 
The reason $c_B$ does much worse with $ggg$ is because the lensing kernel is much more sensitive to $c_B$ than the growth of perturbations, as was also the case for the power spectrum shown in Fig.~\ref{fig:cl}. This is also evident from Fig.~\ref{fig:bis_signal_nzbin1_cB}, when comparing for example the fractional deviation curves for the  $\kappa_{(1)}\kappa_{(1)}\kappa_{(1)}$ and $g_{(1)}g_{(1)}g_{(1)}$ bispectra in the second and third panels.

The total result combining both kinds of probes (black solid line) gives an improvement a factor of (1.6, 1.5, 1.5) over the $\kkk$ only results for the ($c_M$, $c_B$, $\Omega_m$) constraints, and a factor of (8, 52, 3) over $ggg$ alone for the same parameters. This provides motivation for combining both the lensing convergence and galaxy density probes in a bispectrum analysis for Horndeski models.

Furthermore, we see that the use of cross-bispectra is important for obtaining such results. While the $\kkk$ constraints are dominated by its subset of auto-bispectra (dashed blue line), the total result is not dominated by simply combining all the auto-bispectra (dashed black line) where
$\mathcal{M} = \{ M = 
\kappa_{(i)} \kappa_{(i)} \kappa_{(i)} \; \mathrm{or}\; g_{(i)}g_{(i)}g_{(i)} \; | \; M \in \mathcal{M}^{\rm tot} \}$. 
(We did not show auto-bispectra for $ggg$, since this is already a set of auto-bispectra by definition, see section~\ref{sec:unique_observables}.)
It is the inclusion of all the cross-bispectra between $\kappa(i)$ and $g(j)$ that contributes to improving constraints over auto-bispectra alone -- a factor of (1.3, 1.1, 1.3) improvement on the ($c_M$, $c_B$, $\Omega_m$) errors respectively. 

We also list in Table~\ref{tab:results} the 1D marginalized constraints at 68\% confidence level for the total bispectrum results for various parameters. In particular, we have $\sigma_{c_M} = 1.8$ and $\sigma_{c_B} = 0.65$ for $n_{\rm zbin} = 3$. 

\begin{figure*}[ht]
\includegraphics[width=0.95\textwidth]{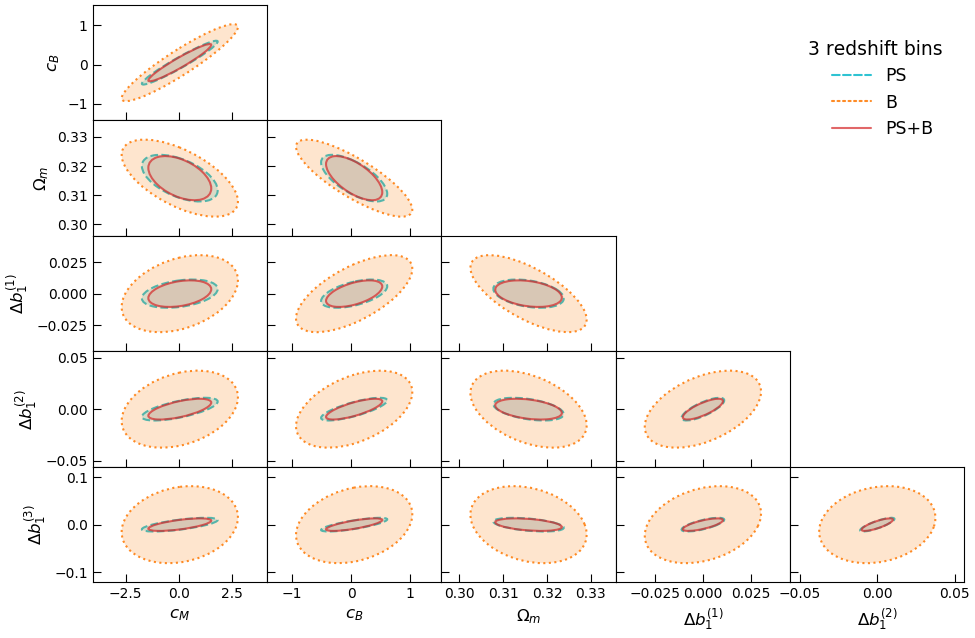}
\caption{Marginalized 2D parameter constraints at 68$\%$ confidence level from the power spectra alone (PS, green dashed), the bispectra alone (B, orange dotted) and their combination (PS+B, red solid) for the fiducial $l_{\rm max} = 1000$ and $n_{\rm zbin} = 3$ setup. The combined PS~+~B constraints improve on power spectra alone by about a factor of $\sim$1.2 for both $c_M$ and $c_B$.}
\label{fig:fisher_nzbin3_lmax1000}
\end{figure*}

\subsection{Combined power spectrum and bispectrum results}
\label{sec:total_results}

We now proceed to combining the power spectrum and bispectrum results. We show in Fig.~\ref{fig:fisher_nzbin3_lmax1000} the power spectrum (PS) constraints in green dashed, the bispectrum (B) in orange dotted and the combined PS + B in red solid, here again for our fiducial choice of $l_{\rm max} = 1000$ and $n_{\rm zbin} = 3$. The bispectrum does worse on its own than the power spectrum constraints alone, but adding the bispectra improves the constraints on both MG parameters by a factor of about $\sim$1.2 compared to PS alone, with modest improvement (about 1.1) for the other parameters without priors ($\{ \Omega_m, b_{1,(i)}\} $). The same kind of improvement is observed for $n_{\rm zbin} = 2$ (not shown here).

The degeneracy directions in the $c_M-c_B$ plane for the bispectra and power spectra are surprisingly similar. They are however more visibly different for the bias parameters. This makes sense since the power spectra and bispectra are proportional to different powers of the galaxy bias. It seems that the improvement in MG parameters mostly comes from the breaking of degeneracy in the bias planes, and that better bispectrum constraints in those planes would lead to improved MG parameter constraints as well.

\begin{table}[h!]
\caption{Marginalized 1$\sigma$ parameter constraints from power spectra (P), bispectra (B) and combined power spectra and bispectra (P+B) for notional survey of 2000 deg$^2$ of the Roman Space Telescope survey overlapped with LSST. We included Planck 2018 priors on the parameters \{$A_s$, $n_s$, $\Omega_b h^2$, $h$\}.   
The second-order biases $b_{2,(i)}$ (not shown here) are also marginalized over for the bispectrum constraints.   
Improvements of the combined results over power spectra alone are about a factor of $\sim$1.3 for $n_{\rm zbin} = 1$, and a about factor of $\sim$1.2 for $n_{\rm zbin} = 2$ and 3 in both parameters $c_M$ and $c_B$.
} 
\label{tab:results}
\begin{center}
\begin{tabular}{l ccc ccc ccc}\hline\hline
&\multicolumn{3}{c}{\hspace{0.5em}no tomography} 
&\multicolumn{3}{c}{\hspace{+0em} 2 redshift bins}
&\multicolumn{3}{c}{\hspace{+1em}3 redshift bins}\\
 & P & B &P+B & P & B &P+B &P &B &P+B\\ \hline
$\sigma(c_M)$ & 2.3 & 3.0 & 1.8 & 1.4 & 2.3 & 1.2 & 1.2 & 1.8 & 0.98\\
$\sigma(c_B)$ & 0.79 & 1.1 & 0.63 & 0.46 & 0.85 & 0.40 & 0.37 & 0.65 & 0.31\\
$\sigma(\Omega_m)$ & 0.009 & 0.016 & 0.008 & 0.006 & 0.011 & 0.006 & 0.005 & 0.009 & 0.005\\
$b_{1,(1)}$ & 0.017 & 0.045 & 0.015 & 0.009 & 0.024 & 0.008 & 0.007 & 0.020 & 0.007\\
$b_{1,(2)}$ & - & - & - & 0.011 & 0.051 & 0.010 & 0.007 & 0.025 & 0.007\\
$b_{1,(3)}$ & - & - & - & - & - & - & 0.010 & 0.054 & 0.009\\
\hline
\end{tabular}
\end{center}
\end{table}

\begin{figure*}[ht]
\includegraphics[width=0.95\textwidth]{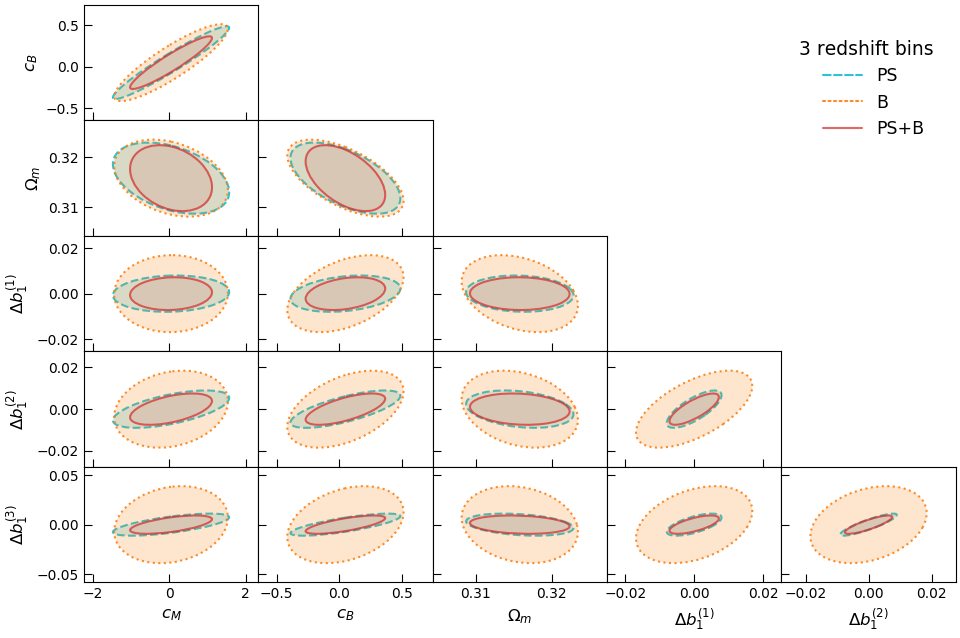}
\caption{Same as Fig.~\ref{fig:fisher_nzbin3_lmax1000} but for $l_{\rm max} = 3000$. Combining the power spectra and bispectra improves on the power alone by about a factor of about $\sim$1.4 for both $c_M$ and $c_B$.}
\label{fig:fisher_nzbin3_lmax3000}
\end{figure*}

In fact, we see in Fig.~\ref{fig:fisher_nzbin3_lmax3000} where we let $l_{\rm max}$ increase from 1000 to 3000, that the bispectrum constraints are closer to those of the power spectrum, because of the much larger number of triangles available at higher multipoles compared to the PS modes. We see there that the improvement for $c_M$ and $c_B$ is also better -- about a factor of $\sim$1.4, while for the rest of the cosmological parameters without priors, it is about a factor of $\sim$ $1.1-1.2$.

\begin{figure}[ht]
\includegraphics[width=0.5\textwidth]{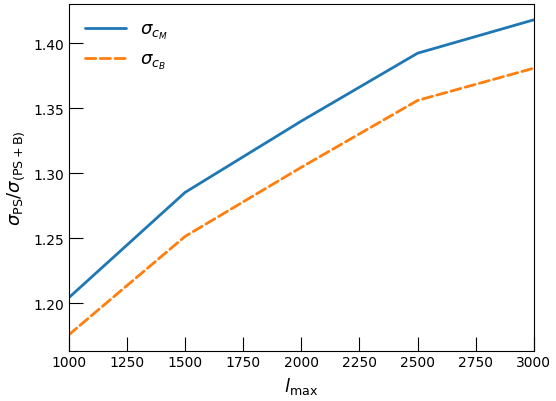}
\caption{Improvement on $\sigma_{c_M}$ and $\sigma_{c_B}$ from adding bispectra to power spectra as a function of the maximum multipole $l_{\rm max}$ for the $n_{\rm zbin} = 3$ case. Results are similar for other $n_{\rm zbin}$.}
\label{fig:improvement_nzbin3}
\end{figure}

To see the trend more clearly, we plot in Fig.~\ref{fig:improvement_nzbin3} the improvement from adding the bispectra on $\sigma_{c_M}$ (blue solid) and $\sigma_{c_B}$ (orange dashed) as a function of $l_{\rm max}$. It is clear that the higher the maximum multipole, the better the improvement gets when adding the bispectra. We also note that these forecasts of improvement at higher $l$ are approximate, because of the nonlinear effects that become stronger at small scales that are not modelled here. However, the nonlinear effects would increase the sensitivity of both the bispectra and the power spectra, and whether the bispectra would benefit much more is to be seen. Nevertheless, the improvement due to the more rapidly growing number of modes in the bispectra would still be present.

\begin{figure}[t]
\includegraphics[width=0.43\textwidth]{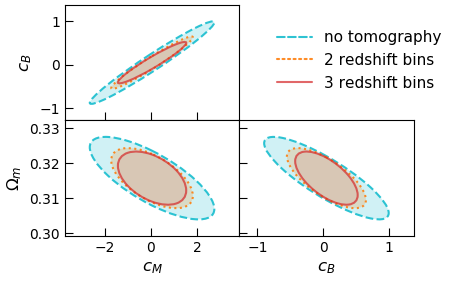}
\caption{Marginalized 2D constraints on parameters from a combined power spectra and bispectra analysis,  varying $n_{\rm zbin} = 1,2$ and 3 at fixed $l_{\rm max} = 1000$. One gains a factor of 1.8 (2.0) improvement in $c_M$ ($c_B$) going from $n_{\rm zbin} = 1$ to 3, and a factor of 1.5 (1.6) from $n_{\rm zbin} = 1$ to 2.}
\label{fig:fisher_vs_nzbin}
\end{figure}

Finally, we show in Fig.~\ref{fig:fisher_vs_nzbin} how the total results vary with different number of redshift bins, for the fiducial choice of $l_{\rm max} = 1000$. It is clear that tomography serves to improve constraints: a factor of 1.5 and 1.6 for $c_M$ and $c_B$ respectively going just from 1 to 2 redshift bins; and a factor of 1.8 and 2.0 for 3 redshift bins. The 1D marginalized constraints for all the parameters without priors for $n_{\rm zbin} = 1, 2$ and 3 can be found in Table~\ref{tab:results}.

From the way the contours shrink, it seems that the improvement will likely be marginal going much beyond $n_{\rm zbin} = 3$. This is consistent with the findings of Ref.~\cite{Rizzato:2018whp}, where the signal-to-noise-ratio of the lensing convergence bispectrum plateaus for $n_{\mathrm{zbin}} \geq 5$ with or without non-Gaussian and super-sample covariance. Given that the lensing bispectrum is main beneficiary of tomography, we expect similar conclusions to hold for our combined bispectrum results, and so do not explore much higher number of redshift bins. Note also that the bispectrum computation becomes expensive for high $n_{\rm zbin}$ as the number of unique tomographic combinations increases quickly with more bins.\\

\section{Summary and Discussion}
\label{sec:conclusion}

In this paper we forecasted the ability of the Roman Space Telescope overlapped with the LSST survey over its notional $2000$ deg$^2$ survey to constrain a sub-class of Horndeski theories, by using galaxy and lensing convergence bispectra in addition to power spectra. In particular, we explored the cross-bispectra as a way to improve constraints over any auto-bispectra alone. We summarize the main results below. They are quoted for our fiducial choice of $l_{\rm max} = 1000$ and three redshift bins unless otherwise stated:

\begin{itemize}
    \item {Combining all possible auto- and cross-bispectra between the two types of probes $\kappa$ and $g$ gave a factor of 1.6 and 1.5 better constraints on $c_M$ and $c_B$ respectively, compared to using $\kkk$ type of bispectra alone, and a factor of 8 and 52 compared to using $ggg$ type alone. 
    }
    \item {Including all possible cross-bispectra between $\kappa$ and $g$ in different tomographic bins contributed to a factor of 1.3 and 1.1 improvement on the $c_M$ and $c_B$ constraints respectively compared to using all the auto-bispectra defined as
    \beq
    \{ \kappa_{(i)}\kappa_{(i)}\kappa_{(i)}, \; g_{(i)}g_{(i)}g_{(i)}\;|\; i = 1, 2 \,..\, n_{\rm zbin} \}. \notag
    \eeq
    }
    \item {Adding the combined bispectrum result to the power spectrum led to a factor of about $\sim$1.2 improvement for both MG parameters, yielding $\sigma_{c_M} = 1.0$ and $\sigma_{c_B} = 0.3$.}
    \item {Varying the $l_{\rm max}$ used, we find that the improvement due to bispectra is increases to about a factor of $\sim$1.4 for both MG parameters for $l_{\rm max} = 3000$. This is primarily due to the greater number of modes in bispectrum with $l_{\rm max}$. While we expect our linear modeling to be less accurate (in fact, more conservative) at increasingly $\l_{\rm max}$ and that the absolute values of the constraints would change with non-linear modeling added, we expect that the relative improvement factor to remain similar.}
    \item {Varying $n_{\rm zbin} = 1, 2$ and 3, we find that using two tomographic bins already gives a factor of 1.5 and 1.6 improvement in the $c_M$ and $c_B$ constraints compared to no tomography; whereas using three bins leads to a factor of 1.8 and 2.0 better constraints. For the Roman + LSST survey considered here, the improvement beyond three redshift bins is likely to be marginal, while it also becomes computationally expensive to go to higher $n_{\rm zbin}$ as the number of bispectra combinations scales rapidly with $n_{\rm zbin}$.}

\end{itemize}

We caution the readers that these results were obtained using modeling that is solely valid in the linear regime, while the observables are integrated along the line-of-sight and in principle capture scales down to the non-linear regime. While we varied one $l_{\rm max}$ as a very crude way to control the impact of nonlinear scales, there exists more refined methods such as using a different $l_{\rm max}$ per redshift bin corresponding to a desired $k_{\rm max}$, e.g.  Ref.~\cite{Alonso:2016suf}. 

Another less explored method but highly relevant for data analysis is to cut out actual physical scales by forming the appropriate linear combinations of the observables, by extending the $k$-cut method originally proposed in Ref.~\cite{Taylor:2018snp} for $C_l$'s.
As the nonlinear modeling of MG theories are still underway (templates have been suggested and partially tested on equilateral triangles for a few modified gravity models in Ref.~\cite{Bose:2019wuz}), a $k$-cut method for bispectra could help to control the exact $k_{\rm max}$ allowed by the modeling available at the time of data analysis.

Regardless of the caveats named above, we expect that the general observation that cross-bispectra could be powerful at breaking parameter degeneracy to remain applicable to many cases and extendable to other experiments as well. Therefore, this work opens the way for combining multiple probes in higher-order statistics, and providing more avenues for maximizing the information content of next-generation large-scale-structure surveys. 

\begin{acknowledgements}
We thank Masahiro Takada, Bhuvnesh Jain, Wayne Hu, Tim Eifler, Atsushi Taruya, Ben Bose, Hiroyuki Tashiro, Hayato Motohashi, Zachary Slepian, Kris Pardo, and Agnes Fert\'e for useful discussions. We thank the Nancy Grace Roman Space Telescope \emph{Cosmology with the High Latitude Survey} Science Investigation Team for providing feedback for this work and the redshift distributions used in the forecast. C.H. especially thanks Miguel Zumalacárregui for providing private versions of \texttt{hi\_class} and guidance on using the software; C.H. also thanks Tessa Baker for sharing her personal notes on Horndeski theories. Part of this work was done at Jet Propulsion Laboratory, California Institute of Technology, under a contract with the National Aeronautics and Space Administration. Copyright 2020. All rights reserved. 
\end{acknowledgements}

\appendix

\section{Second-order expansion of Wigner-3j symbols with Stirling approximation}
\label{sec:wigner}

\begin{figure}[t]
\includegraphics[width=0.48\textwidth]{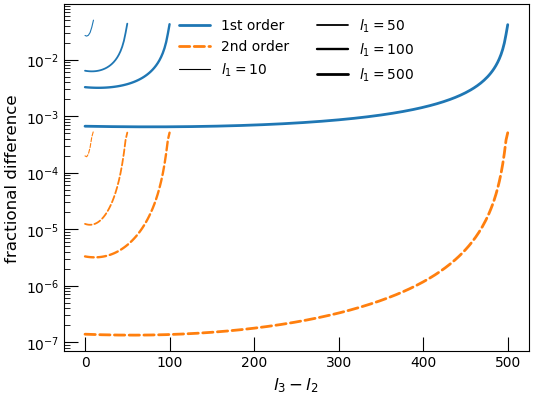}
\caption{Fractional difference between the Stirling approximation and the exact computation of the Wigner-3j symbols as a function of $l_3 - l_2$ for fixed $l_1 = l_2$. The commonly used first order approximation (solid lines) is known to reach sub-percent level accuracy on scales of interest $l\geq 50$ except for degenerate triangles (e.g. Ref.~\cite{Takada:2003ef}). The second order approximation (dashed lines), on the other hand, requires negligible additional computation but reduces the error on degenerate triangles by more than one order of magnitude, reaching sub-percent level accuracy for all configurations.
}
\label{fig:wigner}
\end{figure}

Calculating the bispectrum involves evaluating the
Wigner-3$j$ symbol, which has a closed algebraic form (e.g., Hu 2000):

\bea
\left(
\begin{array}{ccc}
l_1&l_2&l_3 \\
0&0&0
\end{array}
\right)&=&(-1)^{L}\frac{L!}{\left(L-l_1\right)!
\left(L-l_2\right)!\left(L-l_3\right)!} \notag \\
&&\left[\frac{(2L-2l_1)!(2L-2l_2)!(2L-2l_3)!}{(2L+1)!}\right]^{1/2}  \notag\\
\label{eq:w3j}
\eea
for even $l_1+l_2+l_3$ and zero for odd $l_1+l_2+l_3$, where we have also defined $L = (l_1+l_2+l_3)/2$.

Because evaluating the exact expression involves calculating factorials $l!$ which diverges for large $l$, we employ an approximation based on the Stirling approximation:
$n!=\Gamma(n+1)$ and 
\begin{equation}
\Gamma(x)\sim (2\pi)^{1/2}e^{-x}x^{x-1/2},~ ~ \mbox{  for large  }x. 
\end{equation}

While the commonly used first order expansion is good enough with errors less than $0.2\%$
for angular scales $l\ge 50$ of our interest for most triangles, it is known to be less accurate for the degenerate triangles, with errors reaching sometimes above percent level (see Fig.~\ref{fig:wigner} for an illustration). We therefore expand the expression to second order, and found that it reduces the error by more than an order of magnitude for degenerate triangles, rendering it sub-percent. At the same time, the errors on other configurations are in general $< 10^{-5}$. These improvements are obtained with negligible additional computational cost and we recommend using them for any bispectrum calculations. 

The exact expression is given by
\begin{widetext}

\begin{eqnarray}
\left(
\begin{array}{ccc}
l_1&l_2&l_3\\
0&0&0
\end{array}
\right)
\approx
(-1)^L \sqrt{\frac{e^3}{2\pi}} (L+1) ^{-1/4}
\,\left[ \prod_{i=1}^3(L-\l_i+1)^{-1/4}\left(\frac{L-\l_i+1/2}{L-\l_i+1}\right)^{L-\l_i+1/4} \right] F(l_1, l_2, l_3),
\end{eqnarray}
where $F(l_1, l_2, l_3) = 1$ for the commonly used first order expansion, and 
\beq
F(l_1, l_2, l_3) = \left( 1 + \frac{1}{12(L+1)} \right) \left( 1 + \frac{1}{24 (L + 1) }\right)^{-1/2} \prod_{i=1}^3 \left(1 + \frac{1}{12 (L-l_i +1)} \right)^{-1} \left( 1 + \frac{1}{24(L-l_i+1/2)} \right)^{1/2}.
\eeq
for the second order expansion used in this paper.
 
\end{widetext}

\section{The expressions for $\lambda(a)$ in $F_2$ for $\alpha_i(a) = c_i \Omega_{\rm DE}(a)$ Horndeski theories}
\label{sec:lambda}

We model the modified gravity effects in the bispectrum up to second order in perturbation theory, where the second-order kernel $F_2$ is modified with a parameter $\lambda(a)$. In section~\ref{sec:bis_mg} we introduced the ansatz $\lambda(a) \approx \widetilde\Omega_m^\xi(a)$. We now summarize briefly the first order expansion of $\xi$ in terms of $c_i$'s that we used to compute $\lambda(a)$ in this work and we refer the readers to Ref.~\cite{Yamauchi:2017ibz} for more details. 

For $\alpha_M(a) = c_M \left(1-\widetilde\Omega_{\rm m}(a) \right)$, we can expand $\xi$ in orders of $\Omega_{\rm DE}$ where the leading order approximation is given by

\al{
	\xi
		=\frac{-3+6\tilde{\gamma}+2\kappa_\Phi^{(1)}+7\tau_\Phi^{(1)}}{(7-6w^{(0)}+2c_{\rm M})(1-3w^{(0)}+c_{\rm M})}
	\,,
\label{eq:Gamma estimation}
}
where
\al{
	\tilde{\gamma}\approx \frac{3-3w^{(0)}-2\kappa_\Phi^{(1)}}{5-6w^{(0)}+2c_{\rm M}}
	\, , 
\label{eq:gamma estimation}
}
and $w^{(0)}$, $\kappa_\Phi^{(1)}$ and $\tau_\Phi^{(1)}$ are lowest order coefficients in the expansion of the dark energy equation of state $w_{\rm DE} = p_{\rm DE}/\rho_{\rm DE}$, $\kappa_{\Phi}$ and $\tau_{\Phi}$ respectively
\al{
	&w_{\rm DE}=\sum_{n=0}\frac{1}{n!}w^{(n)}\left( 1-\widetilde\Omega_{\rm m}\right)^n
	\,,\\
	&\kappa_\Phi -\frac{3}{2}\widetilde\Omega_{\rm m} =\sum_{n=1}^\infty\frac{1}{n!}\kappa_\Phi^{(n)}\left( 1-\widetilde\Omega_{\rm m}\right)^n\,, \\
	&\tau_\Phi 
	=\sum_{n=1}^\infty\frac{1}{n!}\tau_\Phi^{(n)}\left( 1-\widetilde\Omega_{\rm m}\right)^n
	\,.
\label{eq:kappa_Phi exp}
}

\begin{widetext}
They can be written in terms of $c_i$'s in theories where $\alpha_i(a) = c_i \left(1-\widetilde\Omega_{\rm m}(a) \right)$ as
\al{
	&\kappa_\Phi^{(1)}
		=\frac{3}{2}
			\biggl[
			c_{\rm T}
			+\frac{(c_{\rm B}+2c_{\rm M}-2c_{\rm T})^2}{6(1+w^{(0)})-6c_{\rm B}w^{(0)}+2c_{\rm B}c_{\rm M}-c_{\rm B}+4c_{\rm M}-4c_{\rm T}}
			\biggr]
	\,,
\label{eq:kappa_Phi^1}
}
\al{
	\tau_\Phi^{(1)}=&\frac{1}{3}\left(\kappa_Q^{(0)}\right)^2\biggl[\left(c_{\rm B}+c_{\rm M}-\frac{3}{2}c_{\rm T}\right)\kappa_Q^{(0)}-\frac{9}{4}c_T\biggr]
	\notag\\
	&
					+\frac{1}{3}\left(\kappa_Q^{(0)}\right)^2\biggl[\left( 1+3w^{(0)}-c_{\rm M}\right)\kappa_Q^{(0)}-\frac{9}{2}\biggr]c_{\rm V1}
	\notag\\
	&
					-\frac{1}{2}\kappa_Q^{(0)}\left(\kappa_Q^{(0)}+\frac{3}{2}\right)\biggl[\left(\frac{1}{2}-3w^{(0)}+c_{\rm M}\right)\kappa_Q^{(0)}+3\biggr]c_{\rm V2}
	\,,\label{eq:tau_Phi^1}
}
where
\al{
    &\kappa_Q^{(0)}
		=-\frac{3(c_{\rm B}+2c_{\rm M}-2c_{\rm T})}
		{6(1+w^{(0)})-6c_{\rm B}w^{(0)}+2c_{\rm B}c_{\rm M}-c_{\rm B}+4c_{\rm M}-4c_{\rm T}}
	\,. 
\label{eq:kappa_Q^1}
}
With these expressions $\xi$ can be expressed completely in terms of the constant parameters $\{ w^{(0)}\,, c_{\rm B}\,,c_{\rm M}\,,c_{\rm T}\,,c_{\rm V1}\,,c_{\rm V2}\}$. We assume $\Lambda$CDM cosmology as the background expansion so that $w^{(0)}=-1$ throughout. Note that $c_{\rm V1}$ and $c_{\rm V2}$ are constants of proportionality for the functions $\alpha_{V1}$ and $\alpha_{V2}$. While $\alpha_i, i = M, B, K, T$ describe the first order degrees of freedoms of the Horndeski Lagrangian, $\alpha_{V1}$ and $\alpha_{V2}$ are part of the second order expansion and whose relations to $G_a$ are given by
\al{
	&M^2\alpha_{\rm V1}=-2X\left( G_{4X}+2XG_{4XX}-G_{4\phi}+2H\dot\phi G_{5X}-XG_{5\phi X}+\dot\phi HXG_{5XX}\right)
	\,,\\
	&M^2\alpha_{\rm V2}=2\dot\phi HXG_{5X}
	\,.
\label{eq:alpha_V2}
}
Although they are in principle arbitrary functions, we restrict ourselves to setting $c_{V1} = c_{V2} = 0$ when evaluating $F_2$. 

Note that we have assumed that the  constant $\xi$ ansatz is a good approximation to the models used here. The validity of the approximation is checked explicitly against solving for $\lambda(a)$ numerically for various choices of $c_i$'s in Ref.~\cite{Yamauchi:2017ibz}. For most cases the deviation is less than 10\% while in some cases the constancy is eventually violated at low-redshifts (e.g. $c_B = 0.2, c_T = 0.5$). We expect that for the much smaller values of $c_i$'s considered here, the deviations would be less significant.

\end{widetext}

\bibliography{bis.bib}

\end{document}